\documentclass[12pt,letterpaper]{article}
\usepackage[utf8]{inputenc}

\usepackage{color}
\definecolor{darkblue}{rgb}{0.1,0.1,.7}
\usepackage[colorlinks, linkcolor=darkblue, citecolor=darkblue, urlcolor=darkblue, linktocpage]{hyperref}
\usepackage[]{amsmath}
\usepackage[]{graphicx}
\usepackage[]{latexsym}
\usepackage[utf8]{inputenc}
\usepackage{slashed,graphicx,color,amsmath,amssymb}
\usepackage[mathscr]{eucal}
\usepackage{mathrsfs}
\usepackage[margin=10pt,font=small,labelfont=bf]{caption}
\usepackage{amscd}
\usepackage{bm}
\usepackage{xcolor}
\usepackage{upgreek}
\usepackage[square, comma, sort&compress,numbers]{natbib}
\usepackage[all,cmtip]{xy}
\usepackage[margin=1.7in]{geometry}
\usepackage{cleveref}
\usepackage[color=cyan!30!white,linecolor=red,textsize=footnotesize]{todonotes}
\usepackage{soul}
\numberwithin{equation}{section}
\hypersetup{
	pdfstartview={FitH},    
	pdftitle={Sphere partition functions and cut-off AdS},    
	pdfauthor={Caputa, Datta, Shyam},     
	colorlinks=true,       
	linkcolor=blue,          
	citecolor=blue!59!black! ,        
	filecolor=magenta,      
	urlcolor=blue           
}

\def\ttb{T{\bar{T}}}

\def\bz{{\bar{z}}}

\def\ie{{\it i.e.~}}
\def\eg{{\it e.g.~}}

\def\nn{\nonumber}

\def\l1{{{1-loop}}}

\def\bz{{\bar{z}}}

\def\n1{\Bigg|_{n=1}}

\def\n{{(n)}}

\def\bz{{\bar{z}}}

\def\beq{\begin{equation}}
\def\eeq{\end{equation}}

\def\bea{\begin{eqnarray}}
\def\eea{\end{eqnarray}}
\def\nn{\nonumber}

\def\l1{{\text{1-loop}}}

\def\bz{{\bar{z}}}

\def\n1{\Bigg|_{n=1}}

\def\n{{(n)}}

\def\vev#1{\langle{#1}\rangle}

\def\bz{\bar{z}}

\def\be{\begin{equation}}
\def\ee{\end{equation}}
\def\bal{\begin{array}{l}}
\def\ba#1{\begin{array}{#1}}  
	\def\ea{\end{array}}
\def\bea{\begin{eqnarray}}
\def\eea{\end{eqnarray}}
\def\beas{\begin{eqnarray*}}
	\def\eeas{\end{eqnarray*}}

\def\nn{\\\nonumber}

\def\vev#1{\langle #1 \rangle} 

\def\ttb{{T\bar{T}}}

\def\nn{\nonumber}
\def\bit{\begin{item}}
	\def\eit{\end{item}}
\def\benu{\begin{enumerate}}
	\def\eenu{\end{enumerate}}

\def\bz{\bar{z}}

 \makeatletter
 \g@addto@macro\bfseries{\boldmath}
 \makeatother

\makeatletter
\if@todonotes@disabled

\else

\fi
\makeatother

\begin{document}

\definecolor{tinge}{RGB}{255, 244, 195}
\sethlcolor{tinge}
\setstcolor{red}

\vspace*{-.3in} \thispagestyle{empty}
\begin{flushright}
YITP-19-09 
\end{flushright}
\vspace{.4in} {\Large
\begin{center}
{\LARGE \bf  Sphere partition functions  \& cut-off AdS}
\end{center}}
\vspace{.2in}
\begin{center}
{Pawel Caputa$^1$, Shouvik Datta$^2$ and Vasudev Shyam$^3$}
\\
\vspace{.3in}
\small{
$^1$  \textit{Center for Gravitational Physics, 
	Yukawa Institute for Theoretical Physics,\\  Kyoto University, 
	Kyoto 606-8502, Japan. }\\ \vspace{.1cm}
\begingroup\ttfamily\small
pawel.caputa@yukawa.kyoto-u.ac.jp\par
\endgroup 
\vspace{.2cm}
$^2$ \textit{Mani L.~Bhaumik Institute for Theoretical Physics, Department of Physics \& Astronomy,\\
	 University of California,
	Los Angeles, CA 90095, USA.}
}\\ \vspace{.1cm}
\begingroup\ttfamily\small
shouvik@g.ucla.edu\par
\endgroup 
\vspace{.2cm}
$^3$ \textit{Perimeter Institute for Theoretical Physics\\
31 N.~Caroline St. Waterloo, ON, N2L 2Y5, Canada.}
\\ \vspace{.1cm}
\begingroup\ttfamily\small
vshyam@pitp.ca\par
\endgroup 


\end{center}

\vspace{.5in}

\begin{abstract}
\normalsize{We consider sphere partition functions of $TT$ deformed large $N$ conformal field theories in $d=2,\, 3, \, 4,\, 5$ and $6$ dimensions, computed using the flow equation.  These are shown to non-perturbatively match with bulk computations of $AdS_{d+1}$ with a finite radial cut-off. We then demonstrate how the flow equation can be independently derived from a regularization procedure of defining $TT$ operators through a local Callan-Symanzik equation. Finally, we show that the sphere partition functions, modulo bulk-counterterm contributions, can be reproduced from Wheeler-DeWitt wavefunctions.}
\end{abstract}

\vskip 1cm \hspace{0.7cm}

\newpage

\setcounter{page}{1}

\noindent\rule{\textwidth}{.1pt}\vspace{-1.2cm}
\begingroup
\hypersetup{linkcolor=black}
\tableofcontents
\endgroup
\noindent\rule{\textwidth}{.2pt}

\section{Introduction}

Our understanding of quantum gravity has been dramatically advanced by the AdS/CFT correspondence. In a sense, it provides a precise framework to tackle the gravity path integral by formulating it non-perturbatively in terms of a quantum field theory. As we still grapple with several challenges in black hole physics and cosmology, we require to develop newer tools for calculating various observables and engineer mechanisms to adapt holography to more general settings. 

A question of fundamental importance is how can we formulate quantum gravity with some specified boundary conditions and can holography turn out to be useful in this context. A situation where we are posed with this problem appears is in the evolution of a closed universe, wherein the wavefunction of interest is calculated with fixed boundary conditions \ie of an initial state \cite{Hartle:1983ai}. For specified spatial boundary conditions, there has been some progress from studies of the holographic renormalization group \cite{deBoer:1999tgo,deHaro:2000vlm}. Here, attempts were made to interpret the radial cut-off as a Wilsonian cut-off in field theory whereby integrating out bulk geometry corresponds to integrating out high energy modes of the field theory \cite{Faulkner:2010jy,Heemskerk:2010hk}. Identifying the radial cut-off in the bulk with a short-distance cut-off seems to work more naturally in this vein \cite{lecholrg}. 

Another procedure for implementing such a regularization is to deform the holographic CFT by some irrelevant operator, the scale associated to which is treated as the cut-off. The important question is then to pinpoint this operator in holographic CFTs. Along these lines, over the last couple of years a novel viewpoint has emerged via the $\ttb$ deformation of QFTs. Discovered first in two dimensions, the $\ttb$ operator induces a solvable, irrelevant, double-trace deformation of QFTs \cite{Zamolodchikov:2004ce,Smirnov:2016lqw,Cavaglia:2016oda}. The one-parameter family of theories parametrized by the $\ttb$ coupling has a number of special properties. Deforming an integrable theory by $\ttb$ preserves integrability.  The deformed partition function obeys diffusion-like equations \cite{Cardy:2018sdv} and is modular invariant in a unique sense \cite{Datta:2018thy,Aharony:2018bad}. The finite-size spectrum of this theory is exactly the same as Jackiw-Teitelbohm gravity coupled to the undeformed `matter' \cite{Dubovsky:2018bmo}. 

The $\ttb$ deformation, for large-$c$ CFTs, has been proposed to be holographically dual to  AdS$_3$ with a finite radial cut-off with Dirichlet boundary conditions \cite{McGough:2016lol}. Within the pure gravity sector, this geometric notion does reproduce some characteristics of the deformed CFT for a specific sign of the coupling \cite{Kraus:2018xrn}. Some further tests of this cut-off gravity/$\ttb$ relation include the finite-size spectrum, signal propagation velocities, stress tensor correlators \cite{Aharony:2018vux,Kraus:2018xrn}, entanglement entropy \cite{Donnelly:2018bef,Chen:2018eqk,Gorbenko:2018oov}. Interestingly, among other observations, \cite{McGough:2016lol} noticed that the flow equation for the trace of the energy momentum tensor in $T\bar{T}$ deformed theory can be re-written as the Hamiltonian constraint in bulk gravity theory on $AdS_3$ spacetime. A different version of $AdS_3$ holography in this context has been put forward in \cite{Giveon:2017myj,Giveon:2017nie}. A Lorenz breaking cousin of this deformation and its holographic interpretation has been proposed in \cite{Guica:2017lia}. An application to de-Sitter holography was studied in \cite{Gorbenko:2018oov}.

The holographic construction with a cut-off has also been generalized to higher dimensions in \cite{Taylor:2018xcy,Hartman:2018tkw} (see also \cite{Chang:2018dge} for a supersymmetric generalization to higher dimensions). The guiding principle behind these works was to define holographic $TT$-deformed theories via a flow equation that originates from the Gauss-Codazzi equation \ie the Hamiltonian constraint. At large $N$, this flow can indeed be seen as coming from the deforming operators that are quadratic in the energy-momentum tensor (see also \cite{Cardy:2018sdv} for generalization in the form of $\det T$). In \cite{Taylor:2018xcy,Hartman:2018tkw} it was shown that such a procedure is consistent with finite cut-off holography through agreements of the quasi-local energy, speed of sound as well as simple correlators within the pure gravity sector. It was also shown in \cite{Shyam:2018sro} that for $d=4$ various results of holographic RG, such as the gradient form of the metric beta functions,  are also captured by such irrelevant double trace deformations involving the stress tensor and other suitably defined deforming operators. 

In this work, we study the sphere partition functions of $TT$ deformed CFTs in $d\geq 2$. The sphere partition function $Z_{\mathbb{S}^d}$ plays an important role in a wide variety of aspects. The QFT on the sphere is free from IR divergences and for several supersymmetric theories it has been computed exactly by localization techniques \cite{Marino:2011eh,Fuji:2011km,Pestun:2007rz,Pestun:2016zxk}. These have led to many precision tests of holography and a better understanding of RG flows. For CFTs in even dimensions, it captures the anomalies. When anomalies are absent, for instance in 3 dimensions, $F=- \log |Z_{\mathbb{S}^3}|$ serves as an analogue to the central charge in counting the degrees of freedom; $F$ decreases along the RG flow from the UV to the IR \cite{Jafferis:2011zi,Klebanov:2011gs}. Although the $F$-theorem is different in flavour from the even dimensional analogues (the $c$ and $a$ theorems), a unified formulation can be achieved by considering the sphere free energy. This quantity also has other uses. In odd dimensional CFTs, the finite piece of the sphere free energy $F_{d}$ measures the entanglement entropy across a spherical region $\mathbb{S}^{d-2}$ in flat spacetime $\mathbb{R}^{d-1,1}$. 

The spectrum of $TT$ deformed CFTs can be computed from the flow equation. The flow equation relies on special factorization properties of the $TT$-operator \cite{Zamolodchikov:2004ce} which are expected to hold true for large $N$ theories in higher dimensions. For the case of the cylinder $\mathbb{S}^{d-1}\times \mathbb{R}^1$, this takes the form of the Burgers' equation in hydrodynamics. For the sphere on the other hand, owing to its maximal symmetry, the flow equation can be reduced to an algebraic equation and be solved exactly. This allows us to evaluate the sphere free energy/partition function from the field theory side in a very simple manner. In the cut-off AdS bulk, we calculate the on-shell action with the necessary counterterms \cite{Emparan:1999pm} and observe a precise agreement with the field theory analysis.

Since the holographic flow equation was deduced in \cite{Hartman:2018tkw} from re-writing the bulk Gauss-Codazzi equation in terms of the holographic stress tensor, one may wonder how universal or generic is this flow. Proving this equation starting from the definition of $TT$ operators in higher dimensions and on a curved background in ABJM or $\mathcal{N}=4$ SYM is beyond the scope of this work and remains an important future problem. However, we address the issue of how one can obtain such a universal flow equation from purely field theoretic considerations, namely starting already from a local Callan-Symanzik equation describing a CFT deformed only by an irrelevant $TT$ operator. 

The Wheeler-DeWitt equation is a quantum constraint equation in a theory of quantum gravity that encodes the independence of the theory under choice of a foliation of space-time by co-dimension one hypersurfaces. Such a foliation is typically chosen in order to pass to the Hamiltonian formalism, as introduced  in \cite{ADM}. The diffeomorphism invariance of the gravitational theory translates into the independence under the choice of foliation, and is thereby encoded in a constraint. In our context, the relevant foliation of the bulk is by successive finite radius cut-off surfaces. The $TT$ flow equation is then mapped to the semiclassical limit of the bulk Wheeler-DeWitt equation. In this work, we solve the mini-superspace Wheeler-DeWitt equation in the WKB approximation and find the Wheeler-DeWitt wavefunction that, indeed, up to holographic counterterms, matches our partition functions.

This paper is organized as follows. In section \ref{sec:holography}, we compute holographic stress-tensor and partition function in Euclidean anti-de Sitter geometries with spherical boundary at finite radial cut-off up to 6 dimensions. In section \ref{sec:field-theory}, we review the $TT$ deformation in field theory and consider the flow equation on the sphere and its solution. This allows us to obtain the sphere partition function and find an exact agreement with the gravity results, at large $N$. We provide field theoretical derivation of the flow equation starting from the local Callan--Symanzik equation and the regularization procedure of the $TT$ operator in section \ref{sec:FlowRegularization}. In Section \ref{sec:Wheeler-DeWitt}, we discuss the (mini-superspace) Wheeler-DeWitt equation and its solution in the WKB approximation that captures our partition functions up to holographic counterterms. Finally, we conclude and pose some open problems in section \ref{sec:conclusions}. Appendix \ref{sec:Gauss-Codazzi} contains a review of the Gauss-Codazzi equation as a flow equation. Appendix \ref{sec:generateR} has some details on the field theory derivation of the flow. 

\section{Finite cut-off holography}
\label{sec:holography}
We begin by computing the holographic energy-momentum tensor as well as the sphere partition functions up to $d=6$ dimensions using AdS/CFT with a finite cut-off. In standard holographic computations, at large $N$ and strong 't Hooft coupling, both quantities are related to the bulk gravity action evaluated on Euclidean AdS solution in one higher ($d+1$) dimension\footnote{We ignore additional internal directions  in this work.}. 

More precisely, we consider regularized gravity action given by the Einstein-Hilbert (EH) and Gibbons-Hawking (GH) terms supplemented by local counterterms 
\bea\label{on-shell-init}
I^{(d+1)}_{on-shell}=-\frac{1}{2\kappa^2}\int_{\mathcal{M}}d^{d+1}x\sqrt{g}\left(R-2\Lambda\right)+\frac{1}{\kappa^2}\int_{\partial \mathcal{M}}d^dx \sqrt{\gamma}K+S_{ct},
\eea
where the counterterm action, up to $d=6$, takes the form \cite{Emparan:1999pm}
\be\label{counterterm-action}
S_{ct}=\frac{1}{\kappa^2}\int_{\partial \mathcal{M}}d^dx\sqrt{\gamma}\left[\frac{d-1}{l}c^{(1)}_d+\frac{c^{(2)}_d\,l}{2(d-2)} \tilde{R}+\frac{c^{(3)}_dl^3}{2(d-4)(d-2)^2}\left(\tilde{R}^{ij}\tilde{R}_{ij}-\frac{d}{4(d-1)}\tilde{R}^2\right)\right],
\ee
where $\kappa^2\equiv 8\pi G_N$, $\tilde{R}$ and $\tilde{R}_{ij}$ are the Ricci scalar and Ricci tensor of the cut-off surface. To keep track of different contributions, we introduce $c^{(1)}_d=1$ for $d\ge 2$, $c^{(2)}_d=1$ for $d\ge 3$ and $c^{(3)}_d=1$ is non-zero from $d\ge 5$.

We now consider a Euclidean AdS solution 
\be
ds^2=\frac{l^2\,dr^2}{l^2+r^2}+r^2d\Omega^2_d\equiv\frac{l^2\,dr^2}{l^2+r^2}+\gamma_{ij}(r,x)dx^idx^j, \label{metric}
\ee
such that for a fixed value of $r=r_c$ the induced metric, $\gamma_{ij}(r_c,x)=r^2_c\gamma^b_{ij}(x)$, describes a sphere with radius $r_c$. Metric $\gamma^b_{ij}(x)$ of the unit sphere will later be identified with the metric of the boundary QFT theory.

The full on-shell action corresponding to this solution can be used to compute the energy-momentum tensor (Brown-York) and the holographic sphere partition function 
\bea
T^d_{ij}[r]\equiv-\frac{2}{\sqrt{\gamma}}\frac{\delta I^{(d+1)}_{on-shell}[r]}{\delta \gamma^{ij}},\qquad\qquad \log Z_{\mathbb{S}^{d}}[r]\equiv -I^{(d+1)}_{on-shell}[r].\label{PFT}
\eea
Note that both quantities explicitly depend on the radius $r$ at which we cut off spacetime. In standard holography, we take $r$ to infinity and the counterterm action yields finite answers (modulo logarithmic divergences that correspond to anomalies). However, in the context of finite cut-off holography, we keep the radial dependence finite -- this can be holographically interpreted as a deformation by a generalization of the $T\bar{T}$ operator to arbitrary dimensions. 

In what follows, we evaluate both quantities in \eqref{PFT} from the on-shell action with a finite radial cut-off. From the symmetry of the problem, they are determined by a single function of the radius $\omega[r]$ that we extract from both computations, with exact agreement. We will later demonstrate that this function solves the algebraic flow equation that defines the deformed theory. 

\subsection{Holographic stress-tensors}
The holographic stress-tensor \cite{Balasubramanian:1999re,deHaro:2000vlm} is obtained by variation of the on-shell action with respect to the induced metric on the surface of constant $r=r_c$. We first compute the general variation and then show that, as constrained by the spherical symmetry, for our metric \eqref{metric}, energy momentum tensor is proportional to the metric. Performing the standard variations we obtain \cite{Emparan:1999pm,Balasubramanian:1999re,deHaro:2000vlm}\footnote{Note that the signs of the first two terms in \eqref{hol-stress} differ from equation \cite[eq.~(3.3)]{Hartman:2018tkw} since the extrinsic curvature is defined there with an opposite sign. The Gibbons-Hawking term in \cite[eq.~(A.1)]{Hartman:2018tkw} also appears with a minus sign as opposed to our equation \eqref{on-shell-init} where it appears with a plus sign.}
\begin{align}\label{hol-stress}
T_{ij}=-\frac{1}{\kappa^2}&\left\{K_{ij}-K \gamma_{ij}-c^{(1)}_d\frac{d-1}{l}\gamma_{ij}+\frac{c^{(2)}_d\,l}{(d-2)}\tilde{G}_{ij}\right.\nn\\
&\quad \left.+\frac{c^{(3)}_dl^3}{(d-4)(d-2)^2}\left[2(\tilde{R}_{ikjl}-\frac{1}{4}\gamma_{ij}\tilde{R}_{kl})\tilde{R}^{kl}-\frac{d}{2(d-1)}\left(\tilde{R}_{ij}-\frac{1}{4}\tilde{R}\gamma_{ij}\right)\tilde{R}\right.\right.\nn\\
&\quad -\left.\left.\frac{1}{2(d-1)}\left(\gamma_{ij}\Box \tilde{R}+(d-2)\nabla_i\nabla_j\tilde{R}\right)+\Box \tilde{R}_{ij}\right]\right\},
\end{align}
where $\tilde{G}_{ij}$, $\tilde{R}_{ikjl}$ are the Einstein and Riemann tensors and  $\Box$ is  the Laplace-Beltrami operator of the induced metric $\gamma_{ij}$. On the field theory side, the above holographic stress tensor will be used to construct the operator (or its expectation value) which deforms the CFT. For future reference, we note the contributions of additional terms for $d\geq 3$ in the counterterm action \eqref{counterterm-action}.
Following the conventions of \cite{Hartman:2018tkw} we denote   
\begin{align}\label{C-tensor-def}
  {C}_{ij}=\,&\left\{{c^{(2)}_d}\tilde{G}_{ij}+c^{(3)}_db_d\left[2(\tilde{R}_{ikjl}-\frac{1}{4}\gamma_{ij}\tilde{R}_{kl})\tilde{R}^{kl}-\frac{d}{2(d-1)}\left(\tilde{R}_{ij}-\frac{1}{4}\tilde{R}\gamma_{ij}\right)\tilde{R}\right.\right.
\nn\\
&\quad -\left.\left.\frac{1}{2(d-1)}\left(\gamma_{ij}\Box \tilde{R}+(d-2)\nabla_i\nabla_j\tilde{R}\right)+\Box \tilde{R}_{ij}\right]\right\}.
\end{align}
with 
$b_d=l^2/((d-4)(d-2))$.

Now we evaluate the stress-tensor \eqref{hol-stress} for the metric \eqref{metric} with $r=r_c$, so that the induced metric\footnote{In Section \ref{sec:holography}, $\gamma_{ij}$ refers to this induced metric and we suppress the explicit dependence on $(r_c,x)$. Later, in the field theory part, Section \ref{sec:field-theory}, we will work with boundary $\gamma_{ij}$ related by factor of $r^2_c$. We hope that notation should be clear depending on the context.}  becomes $\gamma_{ij}(r_c,x)$. Firstly, the extrinsic curvature terms on the constant $r=r_c$ surface become
\be
K_{ij}-K  \gamma_{ij}=\frac{d-1}{l}\sqrt{1+\frac{l^2}{r^2_c}} \,  \gamma_{ij}.
\ee
The Ricci tensor at $r_c$ is also proportional to the metric $\tilde{R}_{ij}=\frac{d-1}{r^2_c}\gamma_{ij}$ such that the Einstein tensor for the sphere (Einstein manifold) is given by
\be
\tilde{G}_{ij}=-\frac{(d-2)(d-1)}{2r^2_c}\gamma_{ij}.
\ee
Then, the contraction of the Riemann tensor with Ricci tensor is also proportional to the metric such that the second line of \eqref{hol-stress} becomes
\be
2(\tilde{R}_{ikjl}-\frac{1}{4}\gamma_{ij}\tilde{R}_{kl})\tilde{R}^{kl}-\frac{d}{2(d-1)}\left(\tilde{R}_{ij}-\frac{1}{4}\tilde{R}\gamma_{ij}\right)\tilde{R}=\frac{(d-1)(d-4)(d-2)^2}{8r^4_c}\gamma_{ij}.
\ee
Finally, the last line \eqref{hol-stress} vanishes for the sphere (constant curvature) and we get the holographic energy momentum-tensor \eqref{hol-stress} for the $d$-dimensional sphere at $r=r_c$ in Euclidean $AdS_{d+1}$ \eqref{metric}
\be   \label{omega-from-AdS}
T^d_{ij}[r_c]=\frac{(d-1)}{\kappa^2 l}\left[c^{(1)}_d+\frac{c^{(2)}_dl^2}{2r^2_c}-\frac{c^{(3)}_dl^4}{8r^4_c}-\sqrt{1+\frac{l^2}{r^2_c}}\right]\gamma_{ij}.
\ee
Indeed, we see that it is proportional to the metric and we define the proportionality function as
\be
\omega[r_c]=\frac{(d-1)}{\kappa^2 l}\left[c^{(1)}_d+\frac{c^{(2)}_dl^2}{2r^2_c}-\frac{c^{(3)}_dl^4}{8r^4_c}-\sqrt{1+\frac{l^2}{r^2_c}}\right].
\ee
Clearly, we see that different counterterms in various dimensions contribute with polynomial terms whereas the EH and GH terms yield the square-root part. As we shall show in Section \ref{sec:field-theory}, this function can be obtained by solving the QFT flow equation that becomes an algebraic equation for $\omega[r_c]$.

\subsection{Sphere partition functions}
The next step involves  evaluation of the regularized gravity actions and the holographic sphere partition functions with finite cut-off. We evaluate the action \eqref{on-shell-init} in AdS with a cut-off or wall  at $r=r_c$ where we also take into account the counterterms \eqref{counterterm-action}. 

Our metric \eqref{metric} has a constant negative curvature $R=-d(d+1)/l^2$ and is a solution of the vacuum Einstein equations with negative cosmological constant $\Lambda=-d(d-1)/(2l^2)$. With these ingredients and the formulae of the previous subsection, we can evaluate the on-shell action
\begin{align}
I^{(d+1)}_{\rm on-shell}[r_c]=\frac{dl^{d}S_d}{\kappa^2l}&\left[\int^{r_c}_0\frac{r^d\,dr}{l^{d+1}\sqrt{1+\frac{r^2}{l^2}}}-\left(\frac{r_c}{l}\right)^{d-1}\sqrt{1+\frac{r^2_c}{l^2}}\right.\nn\\
&\quad + \frac{r^d_c}{l^d}\left.\left(c^{(1)}_d\frac{(d-1)}{d}+\frac{c^{(2)}_d(d-1)}{2(d-2)} \frac{l^2}{r^2_c}-\frac{c^{(3)}_d(d-1)}{8(d-4)}\frac{l^4}{r^4_c}\right)\right]
\end{align}
The first term comes from the EH action, the second from the GH boundary term and second line from the counterterms. There is an overall factor of the sphere area, $S_d=(2\pi^{\frac{d+1}{2}})/\Gamma\left(\frac{d+1}{2}\right)$. 
Moreover, the first two terms \ie the EH and GH terms, can be written under one integral as
\be
S_{EH}+S_{GH}=-\frac{d(d-1)S_d}{2\kappa^2l}\int^q_0 \sqrt{l^2q^{d-3}+q^{d-2}}dq,\label{osEHGH}
\ee
where we introduced $q=r^2_c$ and this expression will be important in the Wheeler-DeWitt  analysis (Section \ref{sec:Wheeler-DeWitt}). 
Performing this integral yields the hypergeometric function and writing the answer in terms of $r_c$ gives the full holographic sphere partition function (up to $d=6$)
\begin{align}
\log Z_{\mathbb{S}^{d}}[r_c] =-\frac{dS_dr^d_c}{\kappa^2l}&\left[-\frac{l}{r_c}\,\,_2F_1\left(-\frac{1}{2},\frac{d-1}{2},\frac{d+1}{2},-\frac{r^2_c}{l^2}\right)\right.
 \nn\\
& \quad 
+\left.c^{(1)}_d\frac{(d-1)}{d}+\frac{c^{(2)}_d(d-1)}{2(d-2)}\frac{l^2}{r^2_c}-\frac{c^{(3)}_d(d-1)}{8(d-4)}\frac{l^4}{r^4_c}\right].\label{SPF} 
\end{align}
This is the main result of this section and in Section \ref{sec:Wheeler-DeWitt} we will see how this expression is related to the solution of the Wheeler-DeWitt equation. 

The on-shell action \eqref{SPF}, also allows us to extract $\omega[r_c]$. Namely, in general dimensions, the derivative of the sphere partition function with respect to the radius is related to the expectation value of the trace of the energy-momentum tensor. Therefore, we have 
\be
r_c\partial_{r_c} \log Z_{\mathbb{S}^{d}}[r_c]=-\int d^dx\sqrt{\gamma}\langle T^i_i\rangle=-r^d_cS_d\,d\,\omega[r_c],
\ee
where we used that for the sphere $\vev{T_{ij}}=\omega[r_c] \gamma_{ij}$ and $\vev{T^i_i}=d\,\omega[r_c]$. Differentiating \eqref{SPF} we obtain
\be
\omega[r_c]=\frac{(d-1)}{\kappa^2l}\left[c^{(1)}_d+c^{(2)}_d\frac{l^2}{2r^2_c}-c^{(3)}_d\frac{l^4}{8r^4_c}-\sqrt{1+\frac{l^2}{r^2_c}}\right].
\ee
This is precisely the proportionality function derived in the previous subsection. In the next section, we show that it is the solution of the flow equations (with inclusion of anomalies) in all dimensions that we analyze.

\section{Field theory analysis} 
\label{sec:field-theory}

\subsection{$\ttb$ deformation in general dimensions}
As alluded to in the introduction the $\ttb$ operator was initially introduced in 2$d$ by Zamolodchikov \cite{Zamolodchikov:2004ce}. This bi-local operator is defined as the following quadratic combination of the components of the stress-tensor
\begin{align}\label{tt-def}
\ttb(z,z') = T_{zz}(z) T_{\bz \bz} (z') - T_{z\bz}(z)T_{z\bz}(z').
\end{align}
This definition is in flat Euclidean space ($z=x+it$). By using symmetries and conservation laws of the stress tensor, it can be shown that the expectation value of this operator is a constant. This fact motivates defining the operator at coincident points. Although there are divergences which do appear upon taking the coincident point limit, it can be shown that these appear as total derivative terms. The operator $\ttb$ therefore makes sense unambiguously within an integral. We can then deform a QFT by this operator as follows
\begin{align}\label{action-flow}
\frac{dS(\lambda)}{d\lambda} = \int d^2 x \, \ttb(x). 
\end{align}
It is crucial to observe that the stress tensor components appearing in the right hand side of the above equation are that of the action $S(\lambda)$ and, therefore, the deformation is in a sense recursive. This leads to modifications of the action/Lagrangian which are generically non-linear in the coupling $\lambda$, see \eg \cite{Cavaglia:2016oda,Bonelli:2018kik}. For deformations of CFTs, the $\ttb$ coupling is the only new dimensionful scale of the theory. If a single dimensionful scale is present, the following Ward identity for the effective action holds 
\begin{align}\label{eff-Ward-id}
\lambda \frac{dW}{d\lambda} =  -\frac{1}{2} \int d^2 x \, \vev{T_i^i} . 
\end{align}
Combining the equations \eqref{action-flow} and \eqref{eff-Ward-id} leads to the flow equation
\begin{align}\label{ttb-fac}
\vev{T_i^i} = -2 \lambda \vev{\ttb} = -2 \lambda \left(\vev{T_{zz}}\vev{T_{\bz \bz} }- \vev{T_{z\bz}}^2 \right).
\end{align}
If the theory lives on a cylinder $\mathbb{R}\times \mathbb{S}^1$, the second equality of the above equation takes the same form as the inviscid Burgers' equation of hydrodynamics \cite{Smirnov:2016lqw,Cavaglia:2016oda}. 

The above analysis can be generalized for curved spaces and also to higher dimensions. This has been carried out in \cite{Taylor:2018xcy,Hartman:2018tkw}. The strategy there was to make use of the holographic stress tensor and higher dimensional analogues of \eqref{action-flow} and \eqref{eff-Ward-id} to build the deforming operator. 
Although the factorization property \eqref{ttb-fac} is not true in general for curved spaces and $d> 2$, it is still expected to hold for large $N$ theories. 
The deforming operator has the following structure
\begin{align}\label{x-def}
X_d =  \left(T_{ij}+\frac{\alpha_d}{\lambda^{\frac{d-2}{d}}}C_{ij}\right)^2-\frac{1}{d-1}\left(T^{i}_i+\frac{\alpha_d}{\lambda^{\frac{d-2}{d}}}C^i_i\right)^2+\frac{1}{d }\frac{\alpha_d}{\lambda^{\frac{(d-2)}{d}}}\left(\frac{(d-2)}{2}R+C^i_i\right).
\end{align}
The notation $(B_{ij})^2= B_{ij}B^{ij}$ has been used above.
Here, $\alpha_d$ is a dimensionless parameter depending on the degrees of freedom of the theory -- \eg  $\alpha_4 = N/(2^{7/2}\pi)$ for $\mathcal{N}=4$ super-Yang-Mills with an $SU(N)$ gauge group. The last term in \eqref{x-def} vanishes for $d=3, 4$.
 The tensor $C_{ij}$ is the contribution to the holographic stress tensor from additional counterterms in $d\geq 3$, equation \eqref{C-tensor-def}. For the field theory on a sphere \eqref{C-tensor-def} becomes
 \be
C_{ij}=c^{(2)}_dG_{ij}+c^{(3)}_d\frac{2d\alpha_d\lambda^{\frac{2}{d}}}{d-4}\left[2(R_{ikjl}R^{kl}-\frac{1}{4}\gamma_{ij}R_{kl}R^{kl})-\frac{d}{2(d-1)}\left(RR_{ij}-\frac{1}{4}\gamma_{ij}R^2\right)\right],\label{CijQFT}
\ee
where $c^{(n)}_d$ are defined as in Section \ref{sec:holography} (see below equation \eqref{counterterm-action}) so the first term only appears from $3$ dimensions and the second from $5$ dimensions.\\ 
For even dimensions, the appropriate anomaly terms are included as a part of the deforming operator.\\ 
Even though parameters $\alpha_d$ and $\lambda$ are conveniently introduced for higher dimensional analysis, we can also reproduce the $d=2$ flow equation. This requires some care and we have  (see also appendix A)
\be
\alpha_2=\lim_{d\to 2}\frac{l}{(d-2)\kappa^2}=\lim_{d\to 2}\frac{c}{12\pi(d-2)},
\ee
such that factors of $(d-2)$ in front of $R$ in \eqref{x-def} and in $\alpha_2$ cancel each other and we recover the $T\bar{T}$ flow equation (see below).

 The above operator is quadratic in the stress tensor and should be viewed as the large $N$ approximation of a more general operator which could give rise to the dual quantum field theory for cut-off AdS. Specifically, the deforming operators across various dimensions are given by
\begin{align}
X_2 &= \left( T_{ij}   \right)^2 -  \left( T^i_{i}  \right)^2 + \frac{1}{2\lambda}  \frac{c}{24\pi} R   ,\\
X_3 &= \left( T_{ij} + \frac{\alpha_3}{\lambda^{1\over 3}} G_{ij} \right)^2 - \frac{1}{2}\left( T^i_{i} + \frac{\alpha_3}{\lambda^{1\over 3}} G^i_{i} \right)^2  ,\\
\label{4d-deform}
X_4 &= \left( T_{ij} + \frac{\alpha_4}{\lambda^{1\over 2}} G_{ij} \right)^2 - \frac{1}{3}\left( T^i_{i} + \frac{\alpha_4}{\lambda^{1\over 2}} G^i_{i} \right)^2 ,\\
X_5 &= \left( T_{ij} + \frac{\alpha_5}{\lambda^{3\over 5}} C_{ij} \right)^2 - \frac{1}{4}\left( T^i_{i} + \frac{\alpha_5}{\lambda^{3\over 5}}C^i_{i} \right)^2 +\frac{1}{5\lambda}\frac{\alpha_5}{\lambda^{\frac{3}{5}}}  \left(\frac{3}{2}R + C_i^i\right), \\
X_6 &= \left( T_{ij} + \frac{\alpha_6}{\lambda^{2\over 3}} C_{ij} \right)^2 - \frac{1}{5}\left( T^i_{i} + \frac{\alpha_6}{\lambda^{2\over 3}} C^i_{i} \right)^2  + \frac{1}{6\lambda}\frac{\alpha_6}{\lambda^{\frac{2}{3}}}  \left(2R +C_i^i\right).
\end{align}
In $2d$, the relation $l^2 = \frac{c\lambda}{3\pi}$ has already been used to obtain the form above from \eqref{x-def}.
Also note that in $4d$, the squares appearing \eqref{4d-deform} can be expanded and the terms corresponding to the anomaly can be manifestly separated
\begin{align}
X_4 &=  T_{ij}T^{ij}   - \frac{1}{3}( T^i_{i})^2  + 2 \frac{\alpha_4}{\sqrt{\lambda}} \left( G_{ij}T^{ij} - \frac{1}{3} G_i^i T^i_i \right)  + \frac{1}{4\lambda}\  \frac{C_T}{8\pi} \left(G_{ij}G^{ij} - \frac{1}{3} \left(G_{i}^i\right)^2\right) .
\end{align}
Here we have used the relation between $\alpha_4$ and  the central charge, $C_T=32 \pi \alpha_4^2$ (further  details are provided below). This is the expression for the deforming operator in $4d$ which appears in \cite{Hartman:2018tkw}. Similarly, in 6 dimensions, using
\be
C_{ij}=G_{ij}+6\lambda \left(\frac{\alpha_6}{\lambda^{\frac{2}{3}}}\right)\left[2\left(R_{ikjl}R^{kl}-\frac{1}{4}\gamma_{ij}R_{kl}R^{kl}\right)-\frac{3}{5}\left(RR_{ij}-\frac{1}{4}\gamma_{ij}R^2\right)\right],
\ee
we can write the operator as 
\begin{align}
X_6 =\ &  T_{ij}T^{ij}-\frac{1}{5}( T^i_{i})^2  + 2 \frac{\alpha_6}{\lambda^{2/3}} \left( C_{ij}T^{ij} - \frac{1}{5} C_i^i T^i_i \right) \nonumber\\
 &+\frac{144\alpha^3_6}{6\lambda}\left[R^{ij}R_{ikjl}R^{kl}-\frac{1}{2}RR_{kl}R^{kl}+\frac{3}{50}R^3\right]+\frac{1}{\lambda^{2/3}}O(R^4).
\end{align}
The term with third order in curvature precisely matches the (negative of) the 6$d$ anomaly \cite{Henningson:1998gx} provided
$
\alpha_6={N}/{24\pi}
$. 
Moreover, the terms quartic in curvature can be compactly written as
\be
O(R^4)=(C_{ij}-G_{ij})^2-\frac{1}{5}(2R+C^i_i)^2,
\ee
and they come as important part of the operator needed for the correct solution of the flow equation. 

The operator \eqref{x-def} was arrived at by using the form of the holographic stress tensor \cite{Hartman:2018tkw} (see also Appendix \ref{sec:Gauss-Codazzi}). In section \ref{sec:FlowRegularization}, we will also provide an independent procedure to derive  $X_d$ by using a point-splitting procedure. However, for the rest of this section we assume that this is a 
correct flow equation in large $N$ holographic CFTs and employ in a concrete example.

\subsection{The deformation on $\mathbb{S}^d$}
\label{ssec:remarks}
We now consider the $TT$ deformation of a CFT on the unit sphere $\mathbb{S}^d$. Since the sphere is a maximally symmetric space, the stress-tensor expectation values are proportional to the metric\footnote{Here $\gamma_{ij}$ refers to the metric on a unit sphere and all geometric quantities are computed using this metric.} $\vev{T_{ij}}= \omega_d \gamma_{ij}$. We can solve for $\omega_d$ by using the trace equation in higher dimensions
\begin{align}\label{trace-rel}
\vev{T_i^i}= - d {\lambda}\vev{X} .
\end{align}
Inserting the explicit form of the operators, this equation becomes an algebraic equation for $\omega_d$ which can be compactly written as
\bea\label{flow-alg}
d\,\omega_d&=&d\lambda\left[\frac{d}{d-1}\omega_d^2+\frac{2\alpha_d}{\lambda^{\frac{d-2}{d}}}\frac{1}{d-1}\,C^i_i \omega_d-\frac{1}{d\lambda}\frac{\alpha_d}{\lambda^{\frac{d-2}{d}}}f_d(R)\right]. 
 \eea
 where $C_{ij}$ is defined in \eqref{CijQFT} and the last term only depends on the curvature via
\be
f_d(R)=\left(\frac{d-2}{2}R+C^i_i\right)+d\lambda\frac{\alpha_d}{\lambda^{\frac{d-2}{d}}}\left(C_{ij}C^{ij}-\frac{1}{d-1}(C^i_i)^2\right).
\ee
The quadratic equation \eqref{flow-alg} can be solved for $\omega_d$ in $d=2,3,4,5,6$ and we get a general formula
\be\label{om-general}
\omega_d^{(\pm)}=\frac{d-1}{2d\lambda}\left(1-\frac{2\alpha_d\lambda^{\frac{2}{d}}}{d-1}C^i_i\pm\sqrt{\left(1-\frac{2\alpha_d\lambda^{\frac{2}{d}}}{d-1}C^i_i\right)^2+\frac{4\alpha_d\lambda^{\frac{2}{d}}}{d-1}f_d(R)}\right), 
\ee
where the $-$ sign is taken in order to reproduce the anomalies in even dimensions as $\lambda\to0$. 

In the ``new" holographic dictionary, the $TT$ coupling, $\lambda$, is expressed by the bulk quantities via the relation \cite{McGough:2016lol,Hartman:2018tkw} 
\begin{align}\label{coupling-rel}
\lambda = {4 \pi G_N l \over d r^d_c}. 
\end{align}
We note that this relation implies that the $TT$ coupling is dimensionless. This is because there is an additional rescaling by the radius of the sphere, $r_c^d$. 

Computing the counterterms and using \eqref{coupling-rel}, in all the examples up to $d\le 6$, the above field theory result \eqref{om-general} agrees with the cut-off AdS computation of the  stress tensor \eqref{omega-from-AdS} given $\omega[r_c]=r^{-d}_c\omega_d$ \footnote{This comes form $T^{bulk}_{ij}=r^{2-d}_c T^{bdr}_{ij}$ and our definitions of $\omega$'s.}. We show this explicitly below.

\subsection*{$d=2$}
The case for $d=2$ has been considered earlier in the context of entanglement entropy computations in \cite{Donnelly:2018bef}. We include it here for completeness. For $d=2$ equation \eqref{om-general} is 
\begin{align}\label{d2result}
\omega_2^{(\pm)}=\frac{1}{4\lambda}\left(1\pm\sqrt{1+\frac{c\lambda}{3\pi}}\right).
\end{align}
The solution with a $-$ sign in front of the square-root  agrees precisely with \eqref{omega-from-AdS}, with the identification \eqref{coupling-rel} for $d=2$ and the usual Brown-Henneaux relation $c = \frac{3l}{2G_N}$. The $\lambda\to 0$ limit of the $-$ branch above reproduces the $2d$ trace anomaly appropriately. The $+$ branch is ruled out since it does not reproduce the trace anomaly in the CFT limit.

\subsection*{$d>2$}
For general $d$, the solution \eqref{om-general} of the flow equation on $\mathbb{S}^d$ \eqref{trace-rel}, is given by (with $\vev{T_{ij}}=\omega_d \gamma_{ij}$)
\begin{align}\label{field-theory-expression}
\hspace{-.3cm}\omega_d ^{(\pm)}
=\frac{d-1}{2d\lambda}\left[1+c_{d}^{(2)}\alpha_d\lambda^{\frac{2}{d}}d(d-2)\left(1-c_{d}^{(3)}\frac{\alpha_d\lambda^{\frac{2}{d}}d(d-2)}{2}\right)\pm\sqrt{1+2d(d-2)\alpha_d\lambda^{\frac{2}{d}}}\right]. 
\end{align}  
There are two branches of the solution since the flow equation yields an algebraic equation quadratic in $\omega_d$. 

Now for $3\leq d \leq 6$, the parameter $\alpha_d$ is related to gravitational quantities via the relation\footnote{This can be derived using the relation $a_d r_c^{d-2}=\alpha_d \lambda^{\frac{2-d}{d}}$ and $a_d=\frac{1}{8\pi G_N (d-2)}$ of \cite{Hartman:2018tkw}. Note that \cite{Hartman:2018tkw} works with $l=1$ and therefore powers of $l$ need to be appropriately reinstated.}
\begin{align}\label{alpha-rel}
\alpha_d = \frac{1}{(2d)^{d-2 \over d} (d-2)} \left(l^{d-1} \over 8\pi G_N\right)^{2/d}.
\end{align}
This quantity can be related to the rank of gauge groups of conventional CFT$_d$ duals of $AdS_{d+1}$
as follows
\begin{align}
\alpha_3 =  \frac{N_{\rm ABJM}}{6\, 2^{1/3}\pi^{2/3}}, \qquad 
\alpha_4 = \frac{ N_{\rm SYM}}{2^{7/2}\pi}, \qquad
\alpha_6 = \frac{N_{(2,0)}}{24\pi},
\end{align}
where, we used the relations for the ratio $l^{d-1}/G_N$ for ABJM, $\mathcal{N}=4$ super-Yang-Mills and the 6$d$ (2,0) theory respectively. Moreover, the following relation between $\alpha_d$, $l$ and $\lambda$ can be verified using \eqref{coupling-rel} and \eqref{alpha-rel}
\begin{align}\label{ads-radius}
\frac{l^2}{r^2_c} = 2d(d-2)\alpha_d\lambda^{2/d}. 
\end{align}
Once we use \eqref{ads-radius}, $\omega^{(-)}_d$ is in precise agreement with bulk $\omega[r_c]$ in the bulk stress tensor  \eqref{omega-from-AdS}.

The behavior of $\omega_d^{(+)}$ in the $\lambda\to 0$ limit is divergent and, similar to $2d$, this branch is ruled out since this does not reproduce the trace anomaly appropriately in the CFT limit. The situation here should be contrasted with that of the torus partition function, wherein non-perturbative ambiguities exist for the negative values of the coupling \cite{Aharony:2018bad}. In a sense, the CFT trace anomaly provides an additional constraint for partition functions on the sphere.

Finally, we have added appropriate counterterms to obtain the holographic stress tensor  and while defining the $TT$ operator. Therefore, the $\lambda \to 0$ limit of the deformed $\mathbb{S}^d$ stress tensor  \eqref{field-theory-expression}, for $\omega^{(-)}_d$, is devoid of any divergences even in $d=5,6$.\footnote{These additional counterterms have not been considered  in \cite{Hartman:2018tkw}.} Explicitly, $\omega^{(-)}_d$ has the following forms in the undeformed  CFT limit
\begin{align}
\omega_{3,5}^{(-)} \approx 0, \qquad \omega_4^{(-)} \approx 12\alpha_4^2= \frac{3N_{\rm SYM}^2}{32\pi^2}, \qquad \omega_6^{(-)} \approx -2880\alpha_6^3 = - \frac{5N_{(2,0)}^3}{24\pi^3}. 
\end{align}
These values are perfectly consistent with trace anomalies of the undeformed holographic theory \cite{Henningson:1998gx}.

\section{$TT$ flow equation from the local Callan-Symanzik equation}
\label{sec:FlowRegularization}
The flow equation we have been using so far was derived in \cite{Hartman:2018tkw} starting from the bulk Gauss-Codazzi equation, as explained in appendix \ref{sec:Gauss-Codazzi}, and is taken to be a definition of the dual theory on the boundary. In this section, we shed more light on this flow equation by utilising the Callan-Symanzik (CS) equation for a holographic CFT deformed only by a particular irrelevant operator constructed from the energy momentum tensor\footnote{We would like to stress that, in this section, the Callan-Symanzik equation with only the $TT$ deformation is our starting point and we argue how the full flow equation in curved background emerges from the regularisation procedure of defining the $TT$ operator. We are \textit{not} providing a prescription or an RG scheme that would  justify the use of CS with only $TT$. We would like to thank Edgar Shaghoulian for correspondence and clarifications on this point.}. In this section we will work up to $d=5$ and leave the technicalities of $d=6$ as a future problem.

\subsection{$TT$ flow equation vs local CS equation}
The flow equation at large $N$ that serves as the starting point for the analysis presented in the previous section is
\begin{equation}
T^{i}_{i}+\frac{\alpha_{d}}{d\lambda^{\frac{d-2}{2}}}C^{i}_{i}=-d\lambda\left( \left(T_{ij}+\frac{\alpha_{d}}{\lambda^{\frac{d-2}{2}}}C_{ij}\right)^{2}-\frac{1}{d-1}\left(T^{i}_{i}+\frac{\alpha_{d}}{\lambda^{\frac{d-2}{2}}}C^{i}_{i}\right)^{2}\right)-\frac{(d-2)\alpha_{d}}{2\lambda^{\frac{d-2}{2}}}R. \label{rfe}
\end{equation}
This can be made more compact by introducing the `bare' energy momentum tensor
 \begin{equation}
 \hat{T}^{ij}=T^{ij}+\frac{\alpha_{d}}{\lambda^{\frac{d-2}{2}}}C^{ij},
 \end{equation}
 and now it reads
 \begin{equation}
 \hat{T}^{i}_{i}=-d\lambda \left(\hat{T}^{ij}\hat{T}_{ij}-\frac{1}{d-1}(\hat{T}^{i}_{i})^{2}\right)-\frac{(d-2)\alpha_{d}}{d\lambda^{\frac{d-2}{2}}}R.\label{bfe}
 \end{equation}
In this section, we aim to obtain the above flow equation from a more intrinsically field theoretic starting point. Namely, the local Callan--Symanzik  equation, which expresses the response of the field theory under a local change of scale. This is encoded in the expectation value of the trace of the energy momentum tensor.

First, we notice that on a flat background, the bare flow equation reduces to the one proposed in \cite{Taylor:2018xcy}
\begin{equation}
\hat{T}^{i}_{i}|_{(\gamma_{ij}=\eta_{ij})}=-d\lambda\left(\hat{T}^{ij}\hat{T}_{ij}-\frac{1}{d-1}(\hat{T}^{i}_{i})^{2}\right).
\end{equation}
On such a background, this equation can certainly be seen as coming from the relationship between the energy momentum tensor and the expectation value of a deforming operator\footnote{We assume no other deforming operators are present.}
\begin{equation}
\langle T^{i}_{i}\rangle|_{(\gamma_{ij}=\eta_{ij})}=-d\lambda \langle \mathcal{O}\rangle,
\end{equation}
where $\mathcal{O}(x)$ is the irrelevant operator of interest, and the parameter $\lambda$ is the scale associated to the irrelevant deformation. This relationship is referred to as the local CS equation on flat space.

On curved spaces, this equation generalizes to
\begin{equation}
\langle T^{i}_{i}\rangle=-d\lambda \langle \mathcal{O}(x)\rangle-\mathcal{A}(\gamma). \label{lrg}
\end{equation}
 For our purposes, $\mathcal{A}(\gamma)$ is the holographic anomaly which is present in even dimensions. 
 This equation readily provides the correct flow equation in $d=2$. Here, in the large $c$ limit, we have
\begin{equation}
\langle \mathcal{O}(x)\rangle|_{c\rightarrow \infty}=\lim_{y\rightarrow x}G_{ijkl}(x)\langle T^{ij}(x)T^{kl}(y)\rangle|_{c\rightarrow \infty} =\langle T^{ij} \rangle \langle T_{ij}\rangle -\langle T^{i}_{i}\rangle^{2},
\end{equation}
where $G_{ijkl}=\gamma_{i(k}\gamma_{l)j}-\gamma_{ij}\gamma_{kl}$, and the anomaly takes the form
\begin{equation}
\mathcal{A}(\gamma)=-\frac{c}{24\pi}R(\gamma).
\end{equation}
So, in the end, the two dimensional $T\bar{T}$ deformed flow equation reads
\begin{equation}
T^{i}_{i}=-2\lambda\left( T^{ij}  T_{ij} -(T^{i}_{i})^{2} \right) -\frac{c}{24\pi}R(\gamma),
\end{equation}
where the angle brackets are dropped in the large $c$ limit. From this derivation, we see that the coincidence between conformal anomaly and the Ricci scalar was crucially important. 

This is no longer the case in $d=3,4,5$. In these dimensions, the anomaly in \eqref{lrg} no longer provides for us the Ricci scalar term in \eqref{rfe}. In fact, in $d=3$ and $d=5$ there is no conformal anomaly whilst in $d=4$ the anomaly is quadratic in the curvature. In order to obtain the Ricci scalar term in the flow equation, it must somehow be `generated' from the definition of $\mathcal{O}(x)$. Furthermore, the anomaly in $d=4$ must somehow also be absorbed into the definition of this operator. These issues are addressed in what follows.

\subsection{From local CS equation to the higher dimensional flow equation}
Our aim, as described in the previous section, is to generate the Ricci scalar term in the equation \eqref{rfe}, from the local CS equation \eqref{lrg}.

In dimensions higher than 2, the deforming operator $\mathcal{O}(x)$ is defined as
\begin{equation}
\mathcal{O}(x)=\lim_{y\rightarrow x}\frac{1}{4} \left(T_{ij}(x)-\frac{1}{d-1}T^{k}_{k}(x)g_{ij}(x)\right)T^{ij}(y).
\end{equation}
It will help to introduce 
\begin{equation}
G_{ijkl}(x)=\left(\gamma_{i(k}(x)\gamma_{l)j}(x)-\frac{1}{d-1}\gamma_{ij}(x)\gamma_{kl}(x)\right),
\end{equation}
so that 
\begin{equation}
T_{ij}(x)-\frac{1}{d-1}T^{k}_{k}(x)g_{ij}(x)=G_{ijkl}(x)T^{kl}(x). 
\end{equation}
From the definition of the energy momentum tensor, we have 
\begin{equation}
\langle\mathcal{O}(x)\rangle Z[\gamma]=\lim_{y\rightarrow x}G_{ijkl}(x)\left(\frac{1}{\sqrt{\gamma(x)}}\frac{\delta }{\delta \gamma_{ij}(x)}\left(\frac{1}{\sqrt{\gamma(y)}}\frac{\delta Z[\gamma]}{\delta \gamma_{kl}(y)}\right)\right).
\end{equation}
In order to generate the $R$ term in \eqref{bfe}, we will implement the coincidence limit through the heat kernel. This method is similar to the one of \cite{Ita:2017uvz} although the context is quite different. 
The heat kernel $K(x,y;\epsilon)$, satisfies the property
\begin{equation}
\lim_{\epsilon\rightarrow 0}K(x,y;\epsilon)=\delta(x,y).
\end{equation}
This property should be thought of as an initial condition for the heat equation
\begin{equation}
\partial_{\epsilon}K(x,y;\epsilon)=(\nabla^{2}_{(x)}+\xi R_{(x)})K(x,y;\epsilon).
\end{equation}
We can now implement the point splitting regularization as follows
\begin{equation}
\lim_{y\rightarrow x}G_{ijkl}\langle T^{ij}(x)T^{kl}(y)\rangle Z[\gamma]=\lim_{\epsilon \rightarrow 0}\int \textrm{d}^{d}yK(x,y;\epsilon) G_{ijkl}(x)\frac{1}{\sqrt{\gamma}(x)}\frac{\delta}{\delta \gamma_{ij}(x)} \left(\frac{1}{\sqrt{\gamma}(y)}\frac{\delta Z[\gamma]}{\delta \gamma_{kl}(y)}\right).
\end{equation}
We also exploit the fact that we can add to the effective action terms involving local functions of the metric
\begin{equation}
Z[\gamma]\rightarrow e^{C[\gamma]}Z[\gamma],
\end{equation}
where $C[\gamma]$ is chosen to be 
\begin{equation}
C[\gamma]=\alpha_{0}\left(\epsilon^{\frac{d}{2}-1}\int \textrm{d}^{d}x \sqrt{\gamma}+\frac{(d^{2}-3)\epsilon^{\frac{d}{2}}}{d(d-1)}\int \textrm{d}^{d}x \sqrt{\gamma}R\right).\label{impr}
\end{equation}
Here, $\alpha_{0}$ is a constant given by
\begin{equation}\alpha_{0}=\frac{\alpha_{d}}{\lambda^{\frac{d+2}{2}}}\left(\frac{d-2}{2d^{2}\kappa(d)}\right),
\end{equation}
where 
\begin{equation}
\kappa(d)=\frac{(d^{2}-3)(d(d(9d-11)-28)+42)}{12d (d-1)^{2}}.
\end{equation}
With this choice of $\epsilon$ scaling in the improvement term $C[\gamma]$, one can show (as we do in appendix \ref{sec:generateR}) that the deforming operator becomes
\begin{equation}
\langle \mathcal{O}(x) \rangle= \lim_{\epsilon \rightarrow 0}\int \textrm{d}^{d}y K(x,y,\epsilon)G_{ijkl}(x)\langle T^{ij}(x) T^{kl}(y) \rangle+\alpha_{0}R(x)
\end{equation}
which we then subject to the large $N$ limit to obtain
\begin{align}
\langle \mathcal{O}(x)\rangle|_{N\rightarrow \infty}&=
\lim_{\epsilon \rightarrow 0}\int \textrm{d}^{d}y K(x,y;\epsilon)G_{ijkl}(x)\langle T^{ij}(x) \rangle \langle T^{kl}(y)\rangle+\alpha_{0}R(x) \nn \\
&=G_{ijkl}(x)\langle T^{ij}(x)\rangle \langle T^{kl}(x)\rangle+\alpha_{0}R(x).
\end{align}
Here we have used the fact that the large $N$ factorized two point function does not suffer any coincidence divergences so the limit can be taken to turn the heat kernel into a delta function, and the $y$ integral can be performed. 
 We can plug this back into the local CS equation \eqref{lrg}, which now reads, at large $N$
\begin{equation}
T^{i}_{i}=-d\lambda G_{ijkl}T^{ij}T^{kl}-\frac{(d-2)\alpha_{d}}{2\lambda^{\frac{d-2}{2}}}R-\mathcal{A}(\gamma). 
\end{equation}
In $d=3$ and $d=5$, the anomaly $\mathcal{A}(\gamma)=0$. Here we immediately obtain \eqref{bfe} provided we make the choice $\hat{T}^{ij}=T^{ij}$. 
In $d=4$, the holographic anomaly is given by 
\begin{equation}
\mathcal{A}=-\frac{\alpha^{2}_{4}}{\lambda^{2}}\left(G_{ij}G^{ij}-\frac{1}{3}(G^{i}_{i})^{2}\right),
\end{equation}
where $G_{ij}=R_{ij}-\frac{1}{2}R g_{ij}$ is the Einstein tensor, and $a$ is the anomaly coefficient. This can be absorbed into an improvement of the energy momentum tensor, which is subsumed in the definition of the bare energy momentum tensor
\begin{equation}
\hat{T}^{ij}= T^{ij}+\frac{\alpha_{4}}{\lambda}G^{ij}.
\end{equation}
In other words, the equation
\begin{equation*}
T^{i}_{i}=-4\lambda(T^{ij}T_{ij}-\frac{1}{d-1}(T^{i}_{i})^{2})-\frac{\alpha_{4}}{\lambda}R-\mathcal{A}(\gamma)
\end{equation*}
becomes
\begin{equation}
T^{i}_{i}+\frac{\alpha_{4}}{\lambda}G^{i}_{i}=-4\lambda \left(\left(T^{ij}+\frac{\alpha_{4}}{\lambda}G^{ij}\right)\left(T_{ij}+\frac{\alpha_{4}}{\lambda}G_{ij}\right)-\frac{1}{d-1}\left(T^{i}_{i}+\frac{\alpha_{4}}{\lambda}G^{i}_{i}\right)^{2}\right)-\frac{\alpha_{4}}{\lambda}R,
\end{equation}
hence we get \eqref{rfe}.

\subsection{Limitations of this method}
Despite the promise, we find that in $d=4$, this method allows us to readily obtain \eqref{rfe} where as in $d=3,5$, we automatically obtain \eqref{bfe}. The reason for this distinction is that the absorbing the anomaly into the improvement of the energy momentum tensor occurs only in $d=4$. In $d=3$ and $d=5$, the absence of the anomaly leaves us only with the bare flow equation. The inclusion of the counterterms, especially as involved as in $d=6$, should arise from a further improvement of the energy momentum tensor. 

In other words, the counterterms are accounted for automatically in $d=4$ whereas must be thought of as an additional input in odd dimensions. Perhaps a different method or scheme would directly give us the renormalized flow equation no matter what dimension we are working in, starting from the local CS equation. 

Finally, one can ask what justifies the specific choices such as the powers of $\epsilon$ appearing in the definition of $C[\gamma]$, and the choice of $\xi(d)$ that appears in the appendix \ref{sec:generateR}. For now, we can only offer a post facto justification, in that these choices lead to the form of the flow equation \eqref{bfe}. It would be interesting to find an intrinsically field theoretic justification for this scheme. 

\section{The Wheeler-DeWitt equation}
\label{sec:Wheeler-DeWitt}
In this section, we comment on the role played by the Wheeler-DeWitt equation in deriving the deformed partition function. We shall see that the WKB solution of the (minisuperspace) Wheeler-DeWitt equation perfectly reproduces the bulk and boundary on-shell action without counterterms. 

Let us briefly review the Wheeler-DeWitt equation that arises in the minisuperspace approximation (we closely follow \cite{Caputa:2018asc}). The minisuperspace ansatz for the Euclidean asymptotically $AdS$ metric is defined as
\be
ds^2=\mathbf{N}^2(r)dr^2+a^2(r)d\Omega^2_{d},
\ee
where $\mathbf{N}(r)$ is the lapse function and $a(r)$ is the scale factor.

We first evaluate the EH and GH actions on this metric and then, in the Euclidean gravity  path integral, we redefine the lapse $\mathbf{N}\to \mathbf{N}a^{d-4}$ and introduce a variable\footnote{The main advantage of the $q$ variable here is the canonical kinetic term.} $q=a^2$ such that the action takes the form (see \cite{Caputa:2018asc} and references therein)
\be
S_{EH}+S_{GH}=-\frac{d(d-1)S_d}{2\kappa^2}\int dr \left[\frac{q'^2}{4\mathbf{N}}+\mathbf{N}\left(q^{d-3}+l^{-2}q^{d-2}\right)\right],
\ee
where $S_d$ is the sphere area.

To derive the Hamiltonian we compute the canonical momentum conjugate to $q(r)$
\be
p=\frac{\partial L}{\partial q'}=-\frac{S_d}{\kappa^2}\frac{d(d-1)}{4\mathbf{N}}q',
\ee
and a Legendre's transform yields
\be
H=\mathbf{N}\hat{H}=-\frac{2\kappa^2}{S_d d(d-1)}\mathbf{N}\left[p^2-\left(\frac{d(d-1)S_d}{2\kappa^2l}\right)^2\left(l^2q^{d-3}+q^{d-2}\right)\right].
\ee
Inserting $p=\hbar\frac{d}{dq}$, we derive the Hamiltonian constraint, or the Wheeler-DeWitt equation for the wavefunction $\Psi[q]$ \cite{Caputa:2018asc}
\be
\hat{H}\Psi[q]=\left[\hbar^2\frac{d^2}{dq^2}-\left(\frac{d(d-1)S_d}{2\kappa^2l}\right)^2\left(l^2q^{d-3}+q^{d-2}\right)\right]\Psi[q]=0.
\ee
This equation can be solved exactly in terms of special functions for $d=2,3,4$ (\eg in $d=3$ the solution is the Airy function that reproduces the ABJM partition function \cite{Marino:2011eh,Fuji:2011km} with perturbative $1/N$ corrections). However,  let us focus just on the semi-classical limit, $G_N\to 0$ (large $N$) but with fixed $q$. In this regime we can use the WKB approximation and the leading order solutions are
\begin{align}
\Psi_{\rm WKB}(q) &\approx \exp \left[ \pm  \left(d(d-1)S_d\over 2\kappa^2 l \hbar\right) \int_{0}^{q} \sqrt{ l^2 q^{d-3} + q^{d-2}}\,  dq \right].
\end{align}
Performing the integral, we can see that with $q=r^2_c$, the $-$ sign solution in $d$-dimensions becomes\footnote{We also set $\hbar=1$ at the end.} 
\begin{align}
\Psi_{\rm WKB}[r_c]
=\exp \left[ \frac{dS_dr^{d-1}_c}{\kappa^2 }\,\,_2F_1\left(-\frac{1}{2},\frac{d-1}{2},\frac{d+1}{2},-\frac{r^2_c}{l^2}\right)\right] 
=e^{-\left(I^{\rm on-shell}_{GR}[r_c]-S_{ct}[r_c]\right)}
\end{align}
where we have identified the exponent as the on-shell EH and GH actions (gravity on-shell action without counterterms) evaluated on our Euclidean $AdS$ metric with finite boundary cut-off 
\be
I^{on-shell}_{GR}[r_c]-S_{ct}[r_c]=S_{EH}[r_c]+S_{GH}[r_c],
\ee
computed in \eqref{osEHGH}. Analogous to \cite{Donnelly:2018bef}, this bare partition function (translated to QFT) can be used to compute entanglement entropy and matched with the Ryu-Takayanagi prescription \cite{Ryu:2006bv} applied to a spacetime with finite cut-off. The details of this computation will be presented elsewhere \cite{CaputaHiranoWIP}.

A few comments are in order at this point. Firstly, this concrete example for the sphere illustrates the known fact that the Wheeler-DeWitt wavefunction should be related to the holographic partition function \cite{deBoer:1999tgo,deBoer:2000cz,Papadimitriou:2004ap}. However, it is the minisuperspace approximation that turns this equation into a powerful tool. Secondly, the counterterms (for the full $TT$ partition function that is obtained from the flow equation) are included by additional canonical transformation as explained, for instance, in \cite{lecholrg}. Thirdly, in the large $N$ limit, it is the WKB solution of the Wheeler-DeWitt equation that can be matched with the on-shell action with finite cut-off. It is therefore an interesting future problem to compare the full solution of the Wheeler-DeWitt equation\footnote{As shown in \cite{Caputa:2018asc}, the full WDW wavefunction is an Airy function in 3$d$. In $2d$, the full solution can be written in terms of the $_1F_1$-hypergeometric function, whilst in 4$d$ it is the parabolic cylinder function or a Hermite polynomial upon variable transformations.} with the CFT deformations at finite $N$.

Finally, let us also recall that the connection between solutions to the radial Wheeler-DeWitt equation and the partition function of the $T\bar{T}$ deformed conformal field theories in two dimensions was first noticed in slightly different guise   in \cite{Freidel:2008sh}. The idea there was to define the partition function for the deformed theory through an integral kernel as
\begin{equation}\label{couple}
Z_{\rm QFT}[e]=\int \mathcal{D}f \, e^{\frac{1}{\lambda}\int f^{+}\wedge f^{-}}Z_{\rm CFT}[e+f],
\end{equation}
where $e^{I}_{i}$ is the dyad associated to the metric on the boundary $\gamma_{ij}$. It was then shown that this kernel, when applied to the Weyl Ward identity for the partition function $Z_{\rm CFT}[e]$, resulted in $Z_{QFT}[e]$ satisfying the Wheeler-DeWitt equation. From our discussion above, it follows that this object can be seen as the generating functional for the $T\bar{T}$ deformed theory not including the counter-terms. See \cite{Freidel:2008sh}, \cite{McGough:2016lol} for more details. 

It is intriguing to note that \eqref{couple} is  very similar to the proposal involving coupling the CFT to Jackiw-Teitelboim gravity in \cite{Dubovsky:2018bmo}. It is also a very interesting open problem to find such integral kernels in higher dimensions.

\section{Conclusions}
\label{sec:conclusions}
In this work we further explored generalized $TT$ deformations in large $N$ CFTs and holography with a finite cut-off. We focused on the deformations defined by the trace of the energy-momentum flow equation in holographic CFTs on the sphere. By computing the energy momentum tensor and sphere partition functions holographically (up to $d=6$), we saw that the crucial information is contained in the proportionality function, $\omega[r_c]$, of the stress tensor, $\vev{T_{ij}}=\omega[r_c]\gamma_{ij}$. In the field theory side, this function solves the (algebraic) $TT$ flow equation provided all the non-trivial ingredients of the holographic dictionary like precise anomalies on $\mathbb{S}^d$ as well as relation between the deformation coupling and the gravity parameters. This program can be generalized to other asymptotic geometries as well as black hole solutions and we leave this for future work.

Since the higher dimensional flow equation originates from the Gauss-Codazzi equation, or the Hamiltonian constraint in gravity, the above results may be seen as a consistency check of AdS/CFT. On the other hand, without the $T\bar{T}$ story, the relation between the radial direction and deformation by irrelevant operators would have remained elusive. This is why there is still a lot to be learned about this new ingredient of holography, especially in higher dimensions. In particular, the definition of $TT$ operators on curved manifolds or purely field theory origin of the flow equation remains challenging. In section \ref{sec:FlowRegularization}, we made some progress on the latter and showed how the field theory flow equation emerges from the regularization procedure in defining the $TT$ operator at large $N$. 

We hope that our arguments can be sharpened so that they capture, in arbitrary dimensions, the $1/N$ corrections and additional matter content of the theories. Along these lines, a potentially promising direction to pursue would be to   obtain the flow equation for a holographic theory away from large $N$. This is possible by first upgrading the parameter $\lambda$ to a local function of space, (\ie a source) and then to apply the methods of the local renormalization group in the presence of irrelevant operator deformations as was studied recently in \cite{vanRees:2011ir,Schwimmer:2019efk}. Then, setting the parameter to be constant would lead to a flow equation of the kind we are interested in. 

The Wheeler-DeWitt equation is ubiquitous in quantum gravity and plays an important role in holographic RG \cite{deBoer:1999tgo,deBoer:2000cz,Papadimitriou:2004ap}. In our example we can see that its mini-superspace version can be employed to reproduce the holographic partition function with a finite cut-off. We may hope that the Wheeler-DeWitt equation can guide us in defining the $TT$ operator and identify its expectation value in the flow equation beyond large $N$. In particular, understanding the relation between the coupling of the $TT$ operator and the cut-off in quantum gravity remains a challenge.

Finally, it is important to explore physical quantities under the $TT$ deformation in various dimensions. In particular, how do correlation functions and transport coefficients (\eg $\eta/s$) get modified.  The quasi-normal modes get shifted upon putting a finite cut-off. This should in turn affect the retarded Green's functions. Similarly, an interesting avenue to explore is how thermalization timescales get affected by $TT$. Since the deformation introduces new interaction terms in the Lagrangian and also leads to superluminal signal propagation, one might expect thermalization to occur faster.  Last but not least, many recently developed quantum information theoretic quantities in holography correspond to bulk objects that are non-trivially modified by finite cut-off. Non-perturbative comparisons with deformed CFTs, perhaps even beyond the planar limit, may provide important lessons in this directions (see \eg \cite{Akhavan:2018wla,Hashemi:2019xeq}).

\section*{Acknowledgements}
It is a pleasure to thank John Cardy, Thomas Dumitrescu,  Monica Guica, Michael Gutperle, Shinji Hirano, Mukund Rangamani, Tadashi Takayanagi, Yunfeng Jiang, Per Kraus, Silviu Pufu and Edgar Shaghoulian for fruitful discussions. Some calculations in this work were performed with the aid of the collection of xAct Mathematica packages. PC is supported by the Simons Foundation through the ``It from Qubit" collaboration and by JSPS Grant-in-Aid for Research Activity start-up 17H06787. SD would like to thank the participants and organizers of CHORD`18 at the KITP for simulating discussions on related topics and UC Berkeley and IIT Kanpur for hospitality during the completion of this work. This research was supported in part by the National Science Foundation under Grant No.~NSF PHY-1748958.  VS is supported by the Perimeter Institute for Theoretical Physics. Research at Perimeter Institute is supported by the Government of Canada through Industry Canada and by the Province of Ontario through the Ministry of Research and Innovation.

\appendix
\section{Gauss-Codazzi equation}
\label{sec:Gauss-Codazzi}
In this appendix we give a lightning review of the ideas presented in \cite{Taylor:2018xcy,Hartman:2018tkw} for deriving a flow equation for the trace of the energy momentum tensor in large $N$ holographic CFTs in $d$-dimensions. The main objective of these works is to generalize the $T\bar{T}$ deformations from $2d$ CFTs that result in the finite cut-off in dual holographic geometry (as in \cite{McGough:2016lol}). As observed in \cite{McGough:2016lol}, in two dimensions, the $T\bar{T}$ flow equation can be rewritten as the Gauss-Codazzi equation (Hamiltonian constraint) in gravity. Therefore, the logic to derive analogues in higher dimensions is to derive a flow starting from Gauss-Codazzi equation and postulate that it should be realized as a flow in holographic CFT at large $N$ deformed by the $TT$ operator.

In the Hamiltonian approach to holographic renormalization (see e.g. \cite{Papadimitriou:2004ap,lecholrg}), it is convenient to write Einstein's equations in terms of intrinsic ($\tilde{R}$) and extrinsic ($K$) curvatures of the hypersurfaces $\Sigma_r$ of constant radial direction $r$ with induced metric $\gamma_{ij}$. In this formalism, Einstein's equations are equivalent to Gauss-Codazzi equations. In particular, in the case of pure gravity, their $(r,r)$ component for $d+1$-dimensional spacetime with $d$-dimensional $\Sigma_r$ is given by
\be
K^2-K_{ij}K^{ij}=\tilde{R}+\frac{d(d-1)}{l^2}.\label{GC}
\ee
This equation is just the Hamiltonian constraint namely, with the canonical momentum conjugate to the boundary metric
\be
\pi_{ij}=\frac{\sqrt{\gamma}}{\kappa^2}\left(K\gamma_{ij}-K_{ij}\right),\qquad \pi^i_i=\frac{\sqrt{\gamma}}{\kappa^2}(d-1)K,
\ee
we have
\be
\pi^{ij}\pi_{ij}-\frac{1}{d-1}(\pi^i_i)^2=\frac{\gamma}{\kappa^4}\left[K_{ij}K^{ij}-K^2\right],
\ee
and the Gauss-Codazzi equation (after multiplying by $\kappa^2/\sqrt{\gamma}$) becomes
\be
\frac{\kappa^2}{\sqrt{\gamma}}\left[\pi^{ij}\pi_{ij}-\frac{1}{d-1}(\pi^i_i)^2\right]+\frac{\sqrt{\gamma}}{\kappa^2}\left[\tilde{R}+\frac{d(d-1)}{l^2}\right]=0.
\ee
This is the standard ADM Hamiltonian constraint $H=0$, introduced in \cite{ADM}. This becomes the Wheeler-DeWitt equation $H\Psi=0$ for the wavefunction $\psi$ after replacing canonical momenta with derivatives w.r.t the metric in the quantum theory.

\subsection{Gauss-Codazzi as a holographic flow equation}
The Gauss-Codazzi equation is also equivalent to the flow equation for the expectation value of the trace of the holographic energy-momentum tensor. To see that, take a general form of the holographic stress tensor\footnote{We will drop the expectation values for simplicity of the notation}
\be
\langle T_{ij}\rangle=-\frac{1}{\kappa^2}\left[K_{ij}-K\gamma_{ij}-\frac{d-1}{l}\gamma_{ij}\right]-a_d C_{ij}
\ee
where $\kappa^2=8\pi G_N$ and $a_d=\frac{l}{(d-2)\kappa^2}$ is the known coefficient of the first counterterm above $d=2$. 
It is convenient to introduce a ``bare" stress-tensor
\be
\hat{T}_{ij}=T_{ij}+a_d C_{ij},\label{BareST}
\ee
such that 
\be 
\hat{T}_{ij}=\frac{1}{\kappa^2}\left[K\gamma_{ij}-K_{ij}+\frac{d-1}{l}\gamma_{ij}\right],\qquad \hat{T}^i_{i}=\frac{d-1}{\kappa^2}\left[\frac{d}{l}+K\right].
\ee
From these relations, we have
\be
K^2-K_{ij}K^{ij}=-\frac{2\kappa^2}{l}\hat{T}^i_{i}-\kappa^4\left[\hat{T}_{ij}\hat{T}^{ij}-\frac{1}{d-1}\left(\hat{T}^i_i\right)^2\right]+\frac{d(d-1)}{l^2},
\ee
and again, replacing the LHS with the Gauss-Codazzi equation yields the holographic ``flow" equation for the bare stress-tensor
\be
\hat{T}^i_{i}=-\frac{l\kappa^2}{2}\left[\hat{T}_{ij}\hat{T}^{ij}-\frac{1}{d-1}\left(\hat{T}^i_i\right)^2\right]-\frac{l}{2\kappa^2}\tilde{R}.\label{bareflow}
\ee
We can now also write this equation in terms of the ``renormalized" stress tensor by inserting \eqref{BareST} such that we get \cite{Hartman:2018tkw}
\be
T^i_{i}=-\frac{l\kappa^2}{2}\left[(T_{ij}+a_d C_{ij})^2-\frac{1}{d-1}\left(T^i_{i}+a_d C^i_{i}\right)^2+\frac{4}{\kappa^4}\left(\frac{\tilde{R}}{4}+\frac{a_d\kappa^2}{2l} C^i_{i}\right)\right].\label{FLGR}
\ee
Note that this is still purely phrased in terms of gravitational quantities but we can use the holographic dictionary to turn it into a flow in dual CFTs deformed by a generalized $TT$ operator \cite{Taylor:2018xcy,Hartman:2018tkw}.

\subsection{A dual flow in deformed holographic CFTs}
Formally, in holographic large $N$ CFTs, we can translate the flow equation \eqref{FLGR} to the boundary theory on a unit sphere $\mathbb{S}^d$ by introducing boundary quantities related by powers of the bulk radial cut-off $r_c$. The bulk quantities are translated into the boundary (with superscript $^b$) as \cite{Hartman:2018tkw}
\be
\gamma_{ij}\to r^{2}_c\gamma^b_{ij},\quad T_{ij}\to r^{2-d}_cT^b_{ij} \qquad T^{i}_i=\gamma^{ij}T_{ij}\to r^{-d}_c(T^b)^i_i,\qquad \tilde{R}=r^{-2}_c\tilde{R}^b,
\ee
such that the bare flow equation in QFT becomes
\be
 (\hat{T}^b)^i_i=-\frac{l\kappa^2}{2r^d_c}\left[\hat{T}^b_{ij}(\hat{T}^b)^{ij}-\frac{1}{d-1}\left((\hat{T}^b)^i_i\right)^2+\left(\frac{l\kappa^2}{2r^d_c}\right)^{-1}\frac{lr^{d-2}_c}{2\kappa^2}\tilde{R}^b\right].
 \ee
If we want to interpret this as a QFT flow, we should have
\be
\int d^dx\sqrt{\gamma}\langle (T^b)^i_{i}\rangle=-d\lambda \langle X\rangle,
\ee
and we need in total two relations to replace $l$ and $G_N$ with boundary data. We can write
\be
\lambda=\frac{l\kappa^2}{2dr^d_c},\qquad\qquad \frac{\alpha_d}{\lambda^{\frac{d-2}{d}}}\equiv \frac{lr^{d-2}_c}{(d-2)\kappa^2},\label{Param}
\ee
where $\alpha_d$ is a QFT parameter (see the main text). Using these, we can then write the bare QFT flow as
\be
 (\hat{T}^b)^i_i=-d\lambda\left[\hat{T}^b_{ij}(\hat{T}^b)^{ij}-\frac{1}{d-1}\left((\hat{T}^b)^i_i\right)^2+\frac{1}{d\lambda}\frac{\alpha_d}{\lambda^{\frac{d-2}{d}}}\frac{d-2}{2}\tilde{R}^b\right],
 \ee
 and similarly for the renormalized stress-tensor\footnote{Using $(\hat{T}^b)_{ij}\to (T^b)_{ij}+a_d r^{d-2}_c C_{ij}$. Note that when translating $C_{ij}$ into field theory, in different dimensions, its different components can have a different scaling with $r_c$. Namely, in our examples, in 4d we have $C_{ij}=G_{ij}=G^b_{ij}$ but the extra term in $6$ and $7$ dimensions has a scaling $r^{-2}_c$.}
  \be
 (T^b)^i_i=-d\lambda\left[(T^b_{ij}+\frac{\alpha_d}{\lambda^{\frac{d-2}{d}}}C_{ij})^2-\frac{1}{d-1}\left((T^b)^{i}_i+\frac{\alpha_d}{\lambda^{\frac{d-2}{d}}}C^i_i\right)^2+\frac{1}{d\lambda}\frac{\alpha_d}{\lambda^{\frac{d-2}{d}}}\left(\frac{(d-2)}{2}\tilde{R}^b+C^i_i\right)\right].
 \ee
The expression can be expanded further (for simplicity we drop the superscript $b$) and we get the equation used in the main text
\bea
T^i_i&=&-d\lambda\left[T_{ij}T^{ij}-\frac{1}{d-1}(T^i_i)^2+\frac{2\alpha_d}{\lambda^{\frac{d-2}{d}}}\left(T_{ij}C^{ij}-\frac{1}{d-1}T^i_iC^i_i\right)\right.\nn\\
 &+&\left.\left(\frac{\alpha_d}{\lambda^{\frac{d-2}{d}}}\right)^2\left(C_{ij}C^{ij}-\frac{1}{d-1}(C^i_i)^2\right)+\frac{1}{d\lambda}\frac{\alpha_d}{\lambda^{\frac{d-2}{d}}}\left(\frac{(d-2)}{2}\tilde{R}+C^i_i\right)\right].
 \eea
 where (up to 6d) in field theory we have
 \be
C_{ij}=c^{(2)}_dG_{ij}+c^{(3)}_d\frac{2d\alpha_d\lambda^{\frac{2}{d}}}{d-4}\left[2(R_{ikjl}R^{kl}-\frac{1}{4}\gamma_{ij}R_{kl}R^{kl})-\frac{d}{2(d-1)}\left(RR_{ij}-\frac{1}{4}\gamma_{ij}R^2\right)\right],
\ee
where all the ingredients are of those of the unit metric on the sphere $\gamma_{ij}$.\\
One last comment is that, naively, it appears that this formula is wrong for $d=2$ because it would kill the anomaly. However, for consistency we must have 
\be
\alpha_2=\lim_{d\to 2}a_d=\lim_{d\to 2}\frac{l}{(d-2)\kappa^2}=\lim_{d\to 2}\frac{c}{12\pi(d-2)},
\ee
where we used the Brown-Henneaux relation, and this precisely gives the anomaly piece when $C_{ij}=0$ in $d=2$.
\section{Generating $R$ in the flow equation}
\label{sec:generateR}
The specific $\epsilon$ scaling of the coefficients in the expression \eqref{impr} are chosen such that in the limit $\epsilon\rightarrow 0$, the following terms vanish\footnote{Note that the order of limits here is to first take $\epsilon\rightarrow 0$ with $N$ fixed and then taking $N\rightarrow \infty$ at the end.}
\begin{equation}
\lim_{\epsilon \rightarrow 0}C[\gamma]=0=\lim_{\epsilon \rightarrow 0} \frac{\delta C[\gamma]}{\delta \gamma_{ij}}.
\end{equation}
The second functional derivative however will remain finite, provided we smear it against the heat kernel. This means that if we distribute the limit, we have
\begin{align}
&\lim_{\epsilon \rightarrow 0}\int \textrm{d}^{d}y K(x,y,\epsilon)G_{ijkl}(x)\frac{1}{\sqrt{\gamma}(x)}\frac{\delta }{\delta \gamma_{ij}(x)}\left(\frac{1}{\sqrt{\gamma}(y)}\frac{\delta(e^{C[\gamma]}Z[\gamma])}{\delta \gamma_{kl}(y)}\right)\nn \\
=&\lim_{\epsilon \rightarrow 0}\int \textrm{d}^{d}y K(x,y,\epsilon)G_{ijkl}(x)\frac{1}{\sqrt{\gamma}(x)}\frac{\delta }{\delta \gamma_{ij}(x)}\left(\frac{1}{\sqrt{\gamma}(y)}\frac{\delta Z[\gamma]}{\delta \gamma_{kl}(y)}\right)\nn \\
&+\left(\lim_{\epsilon \rightarrow 0}\int \textrm{d}^{d}y K(x,y,\epsilon)G_{ijkl}(x)\frac{1}{\sqrt{\gamma}(x)}\frac{\delta }{\delta \gamma_{ij}(x)}\left(\frac{1}{\sqrt{\gamma}(y)}\frac{\delta C[\gamma]}{\delta \gamma_{kl}(y)}\right)\right)Z[\gamma].
\end{align}
We then take a closer look at the term on the second line of the RHS in the expression above
\begin{align}
&\left(\lim_{\epsilon \rightarrow 0}\int \textrm{d}^{d}y K(x,y,\epsilon)G_{ijkl}(x)\frac{1}{\sqrt{\gamma}(x)}\frac{\delta }{\delta \gamma_{ij}(x)}\left(\frac{1}{\sqrt{\gamma}(y)}\frac{\delta C[\gamma]}{\delta \gamma_{kl}(y)}\right)\right)Z[\gamma],\nn \\
&= -\alpha_{0}\lim_{\epsilon\rightarrow  0}\left(\left(\frac{d(d^{2}-3)}{(d-1)}\right)\frac{\epsilon^{\frac{d}{2}-1}}{4}K(x,x;\epsilon)+\frac{2\epsilon^{\frac{d}{2}}}{d}\left(\nabla^{2}_{(x)}+\xi R_{(x)}\right)K(x,x;\epsilon)\right)Z[\gamma].
\end{align}
Here 
\begin{equation}
\xi=-\left(\frac{3d^{3}-4d^{2}-9d+14}{2d(d-1)}\right).
\end{equation}
Then, the heat equation implies that we can write 
\begin{align}
&\lim_{\epsilon\rightarrow 0}\alpha_{0}\left(\left(\frac{d(d^{2}-3)}{(d-1)}\right)\frac{\epsilon^{\frac{d}{2}-1}}{4}K(x,x;\epsilon)+\frac{2\epsilon^{\frac{d}{2}}}{d}\left(\nabla^{2}_{(x)}+\xi R_{(x)}\right)K(x,x;\epsilon)\right)\nn \\
&=\alpha_{0}\lim_{\epsilon\rightarrow 0}\left(\left(\frac{d(d^{2}-3)}{(d-1)}\right)\frac{\epsilon^{\frac{d}{2}-1}}{4}K(x,x;\epsilon)+\frac{2\epsilon^{\frac{d}{2}}}{d}\partial_{\epsilon}K(x,x;\epsilon)\right)=\alpha_{0}R(x).
\end{align}


\begingroup\begin{thebibliography}{10}
	
	\bibitem{Hartle:1983ai}
	J.~B. Hartle and S.~W. Hawking, {\it {Wave Function of the Universe}},
	\href{http://dx.doi.org/10.1103/PhysRevD.28.2960}{{\sf Phys. Rev.} {\sf {D28}
		}{\sf (1983) }{\sf 2960--2975}}.
	[Adv. Ser. Astrophys. Cosmol.3,174(1987)].
	
	\bibitem{deBoer:1999tgo}
	J.~de~Boer, E.~P. Verlinde, and H.~L. Verlinde, {\it {On the holographic
			renormalization group}},
	\href{http://dx.doi.org/10.1088/1126-6708/2000/08/003}{{\sf JHEP} {\sf {08}
		}{\sf (2000) }{\sf 003}},
	\href{http://arxiv.org/abs/hep-th/9912012}{{\ttfamily arXiv:hep-th/9912012
			[hep-th]}}.
	
	\bibitem{deHaro:2000vlm}
	S.~de~Haro, S.~N. Solodukhin, and K.~Skenderis, {\it {Holographic
			reconstruction of space-time and renormalization in the AdS / CFT
			correspondence}},  \href{http://dx.doi.org/10.1007/s002200100381}{{\sf
			Commun. Math. Phys.} {\sf {217} }{\sf (2001) }{\sf 595--622}},
	\href{http://arxiv.org/abs/hep-th/0002230}{{\ttfamily arXiv:hep-th/0002230
			[hep-th]}}.
	
	\bibitem{Faulkner:2010jy}
	T.~Faulkner, H.~Liu, and M.~Rangamani, {\it {Integrating out geometry:
			Holographic Wilsonian RG and the membrane paradigm}},
	\href{http://dx.doi.org/10.1007/JHEP08(2011)051}{{\sf JHEP} {\sf {08} }{\sf
			(2011) }{\sf 051}},
	\href{http://arxiv.org/abs/1010.4036}{{\ttfamily arXiv:1010.4036 [hep-th]}}.
	
	\bibitem{Heemskerk:2010hk}
	I.~Heemskerk and J.~Polchinski, {\it {Holographic and Wilsonian Renormalization
			Groups}},  \href{http://dx.doi.org/10.1007/JHEP06(2011)031}{{\sf JHEP} {\sf
			{06} }{\sf (2011) }{\sf 031}},
	\href{http://arxiv.org/abs/1010.1264}{{\ttfamily arXiv:1010.1264 [hep-th]}}.
	
	\bibitem{lecholrg}
	I.~Papadimitriou, {\it {Lectures on Holographic Renormalization}},
	\href{http://dx.doi.org/10.1007/978-3-319-31352-8_4}{{\sf Springer Proc. Phys.}
		{\sf {176} }{\sf (2016) }{\sf 131--181}}.
	
	\bibitem{SSLee}
	S.-S. Lee, {\it {Quantum Renormalization Group and Holography}},
	\href{http://dx.doi.org/10.1007/JHEP01(2014)076}{{\sf JHEP} {\sf {01} }{\sf
			(2014) }{\sf 076}},
	\href{http://arxiv.org/abs/1305.3908}{{\ttfamily arXiv:1305.3908 [hep-th]}}.
	
	\bibitem{Zamolodchikov:2004ce}
	A.~B. Zamolodchikov, {\it {Expectation value of composite field T anti-T in
			two-dimensional quantum field theory}},
	\href{http://arxiv.org/abs/hep-th/0401146}{{\ttfamily arXiv:hep-th/0401146
			[hep-th]}}.
	
	\bibitem{Smirnov:2016lqw}
	F.~A. Smirnov and A.~B. Zamolodchikov, {\it {On space of integrable quantum
			field theories}},
	\href{http://dx.doi.org/10.1016/j.nuclphysb.2016.12.014}{{\sf Nucl. Phys.}
		{\sf {B915} }{\sf (2017) }{\sf 363--383}},
	\href{http://arxiv.org/abs/1608.05499}{{\ttfamily arXiv:1608.05499 [hep-th]}}.
	
	\bibitem{Cavaglia:2016oda}
	A.~Cavagli{\`a}, S.~Negro, I.~M. Sz{\'e}cs{\'e}nyi, and R.~Tateo, {\it {$T
			\bar{T}$-deformed 2D Quantum Field Theories}},
	\href{http://dx.doi.org/10.1007/JHEP10(2016)112}{{\sf JHEP} {\sf {10} }{\sf
			(2016) }{\sf 112}},
	\href{http://arxiv.org/abs/1608.05534}{{\ttfamily arXiv:1608.05534 [hep-th]}}.
	
	\bibitem{Cardy:2018sdv}
	J.~Cardy, {\it {The $T\overline T$ deformation of quantum field theory as a
			stochastic process}},
	\href{http://arxiv.org/abs/1801.06895}{{\ttfamily arXiv:1801.06895 [hep-th]}}.
	
	\bibitem{Datta:2018thy}
	S.~Datta and Y.~Jiang, {\it {$T\bar{T}$ deformed partition functions}},
	\href{http://dx.doi.org/10.1007/JHEP08(2018)106}{{\sf JHEP} {\sf {08} }{\sf
			(2018) }{\sf 106}},
	\href{http://arxiv.org/abs/1806.07426}{{\ttfamily arXiv:1806.07426 [hep-th]}}.
	
	\bibitem{Aharony:2018bad}
	O.~Aharony, S.~Datta, A.~Giveon, Y.~Jiang, and D.~Kutasov, {\it {Modular
			invariance and uniqueness of $T\bar{T}$ deformed CFT}},
	\href{http://arxiv.org/abs/1808.02492}{{\ttfamily arXiv:1808.02492 [hep-th]}}.
	
	\bibitem{Dubovsky:2018bmo}
	S.~Dubovsky, V.~Gorbenko, and G.~Hern{\'a}ndez-Chifflet, {\it {$T\bar{T}$
			Partition Function from Topological Gravity}},
	\href{http://arxiv.org/abs/1805.07386}{{\ttfamily arXiv:1805.07386 [hep-th]}}.
	
	\bibitem{McGough:2016lol}
	L.~McGough, M.~Mezei, and H.~Verlinde, {\it {Moving the CFT into the bulk with
			$ T\overline{T} $}},  \href{http://dx.doi.org/10.1007/JHEP04(2018)010}{{\sf
			JHEP} {\sf {04} }{\sf (2018) }{\sf 010}},
	\href{http://arxiv.org/abs/1611.03470}{{\ttfamily arXiv:1611.03470 [hep-th]}}.
	
	\bibitem{Kraus:2018xrn}
	P.~Kraus, J.~Liu, and D.~Marolf, {\it {Cutoff AdS$_{3}$ versus the $
			T\overline{T} $ deformation}},
	\href{http://dx.doi.org/10.1007/JHEP07(2018)027}{{\sf JHEP} {\sf {07} }{\sf
			(2018) }{\sf 027}},
	\href{http://arxiv.org/abs/1801.02714}{{\ttfamily arXiv:1801.02714 [hep-th]}}.
	
	\bibitem{Aharony:2018vux}
	O.~Aharony and T.~Vaknin, {\it {The TT* deformation at large central charge}},
	\href{http://arxiv.org/abs/1803.00100}{{\ttfamily arXiv:1803.00100 [hep-th]}}.
	
	\bibitem{Donnelly:2018bef}
	W.~Donnelly and V.~Shyam, {\it {Entanglement entropy and $T \overline{T}$
			deformation}},
	\href{http://arxiv.org/abs/1806.07444}{{\ttfamily arXiv:1806.07444 [hep-th]}}.
	
	\bibitem{Chen:2018eqk}
	B.~Chen, L.~Chen, and P.-x. Hao, {\it {Entanglement Entropy in
			$T\overline{T}$-Deformed CFT}},
	\href{http://arxiv.org/abs/1807.08293}{{\ttfamily arXiv:1807.08293 [hep-th]}}.
	
	\bibitem{Gorbenko:2018oov} 
         V.~Gorbenko, E.~Silverstein and G.~Torroba,  {\it {dS/dS and $T\bar T$}},
	\href{https://arxiv.org/abs/1811.07965}{{\ttfamily arXiv:1811.07965 [hep-th]}}.
	
	\bibitem{Giveon:2017myj}
	A.~Giveon, N.~Itzhaki, and D.~Kutasov, {\it {A solvable irrelevant deformation
			of AdS$_{3}$/CFT$_{2}$}},
	\href{http://dx.doi.org/10.1007/JHEP12(2017)155}{{\sf JHEP} {\sf {12} }{\sf
			(2017) }{\sf 155}},
	\href{http://arxiv.org/abs/1707.05800}{{\ttfamily arXiv:1707.05800 [hep-th]}}.
	
	\bibitem{Giveon:2017nie}
	A.~Giveon, N.~Itzhaki, and D.~Kutasov, {\it {$ \mathrm{T}\overline{\mathrm{T}}
			$ and LST}},  \href{http://dx.doi.org/10.1007/JHEP07(2017)122}{{\sf JHEP}
		{\sf {07} }{\sf (2017) }{\sf 122}},
	\href{http://arxiv.org/abs/1701.05576}{{\ttfamily arXiv:1701.05576 [hep-th]}}.
	
	\bibitem{Guica:2017lia}
	M.~Guica, {\it {An integrable Lorentz-breaking deformation of two-dimensional
			CFTs}},
	\href{http://arxiv.org/abs/1710.08415}{{\ttfamily arXiv:1710.08415 [hep-th]}}.
	
	\bibitem{Taylor:2018xcy}
	M.~Taylor, {\it {TT deformations in general dimensions}},
	\href{http://arxiv.org/abs/1805.10287}{{\ttfamily arXiv:1805.10287 [hep-th]}}.
	
	\bibitem{Hartman:2018tkw}
	T.~Hartman, J.~Kruthoff, E.~Shaghoulian, and A.~Tajdini, {\it {Holography at
			finite cutoff with a $T^2$ deformation}},
	\href{http://arxiv.org/abs/1807.11401}{{\ttfamily arXiv:1807.11401 [hep-th]}}.
	
	\bibitem{Chang:2018dge} 
        C.~K.~Chang, C.~Ferko and S.~Sethi, {\it {Supersymmetry and $T \overline{T}$ Deformations}},
        \href{http://arxiv.org/abs/1811.01895}{{\ttfamily arXiv:1811.01895 [hep-th]}}.
	
	\bibitem{Shyam:2018sro}
	V.~Shyam, {\it {Finite Cutoff AdS$_{5}$ Holography and the Generalized Gradient
			Flow}},  \href{http://dx.doi.org/10.1007/JHEP12(2018)086}{{\sf JHEP} {\sf
			{12} }{\sf (2018) }{\sf 086}},
	\href{http://arxiv.org/abs/1808.07760}{{\ttfamily arXiv:1808.07760 [hep-th]}}.
	

	\bibitem{Marino:2011eh}
	M.~Marino and P.~Putrov, {\it {ABJM theory as a Fermi gas}},
	\href{http://dx.doi.org/10.1088/1742-5468/2012/03/P03001}{{\sf J. Stat.
			Mech.} {\sf {1203} }{\sf (2012) }{\sf P03001}},
	\href{http://arxiv.org/abs/1110.4066}{{\ttfamily arXiv:1110.4066 [hep-th]}}.
	
	\bibitem{Fuji:2011km}
	H.~Fuji, S.~Hirano, and S.~Moriyama, {\it {Summing Up All Genus Free Energy of
			ABJM Matrix Model}},  \href{http://dx.doi.org/10.1007/JHEP08(2011)001}{{\sf
			JHEP} {\sf {08} }{\sf (2011) }{\sf 001}},
	\href{http://arxiv.org/abs/1106.4631}{{\ttfamily arXiv:1106.4631 [hep-th]}}.
	
	\bibitem{Pestun:2007rz}
	V.~Pestun, {\it {Localization of gauge theory on a four-sphere and
			supersymmetric Wilson loops}},
	\href{http://dx.doi.org/10.1007/s00220-012-1485-0}{{\sf Commun. Math. Phys.}
		{\sf {313} }{\sf (2012) }{\sf 71--129}},
	\href{http://arxiv.org/abs/0712.2824}{{\ttfamily arXiv:0712.2824 [hep-th]}}.
	
	\bibitem{Pestun:2016zxk}
	V.~Pestun {\em et al.}, {\it {Localization techniques in quantum field
			theories}},  \href{http://dx.doi.org/10.1088/1751-8121/aa63c1}{{\sf J. Phys.}
		{\sf {A50} }{\sf no.~44, }{\sf (2017) }{\sf 440301}},
	\href{http://arxiv.org/abs/1608.02952}{{\ttfamily arXiv:1608.02952 [hep-th]}}.
	
	\bibitem{Jafferis:2011zi}
	D.~L. Jafferis, I.~R. Klebanov, S.~S. Pufu, and B.~R. Safdi, {\it {Towards the
			F-Theorem: N=2 Field Theories on the Three-Sphere}},
	\href{http://dx.doi.org/10.1007/JHEP06(2011)102}{{\sf JHEP} {\sf {06} }{\sf
			(2011) }{\sf 102}},
	\href{http://arxiv.org/abs/1103.1181}{{\ttfamily arXiv:1103.1181 [hep-th]}}.
	
	\bibitem{Klebanov:2011gs}
	I.~R. Klebanov, S.~S. Pufu, and B.~R. Safdi, {\it {F-Theorem without
			Supersymmetry}},  \href{http://dx.doi.org/10.1007/JHEP10(2011)038}{{\sf JHEP}
		{\sf {10} }{\sf (2011) }{\sf 038}},
	\href{http://arxiv.org/abs/1105.4598}{{\ttfamily arXiv:1105.4598 [hep-th]}}.
	
	\bibitem{Emparan:1999pm}
	R.~Emparan, C.~V. Johnson, and R.~C. Myers, {\it {Surface terms as counterterms
			in the AdS / CFT correspondence}},
	\href{http://dx.doi.org/10.1103/PhysRevD.60.104001}{{\sf Phys. Rev.} {\sf
			{D60} }{\sf (1999) }{\sf 104001}},
	\href{http://arxiv.org/abs/hep-th/9903238}{{\ttfamily arXiv:hep-th/9903238
			[hep-th]}}.
	
	\bibitem{ADM}
	R.~L. Arnowitt, S.~Deser, and C.~W. Misner, {\it {The Dynamics of general
			relativity}},  \href{http://dx.doi.org/10.1007/s10714-008-0661-1}{{\sf Gen.
			Rel. Grav.} {\sf {40} }{\sf (2008) }{\sf 1997--2027}},
	\href{http://arxiv.org/abs/gr-qc/0405109}{{\ttfamily arXiv:gr-qc/0405109
			[gr-qc]}}.
	
	\bibitem{Balasubramanian:1999re}
	V.~Balasubramanian and P.~Kraus, {\it {A Stress tensor for Anti-de Sitter
			gravity}},  \href{http://dx.doi.org/10.1007/s002200050764}{{\sf Commun. Math.
			Phys.} {\sf {208} }{\sf (1999) }{\sf 413--428}},
	\href{http://arxiv.org/abs/hep-th/9902121}{{\ttfamily arXiv:hep-th/9902121
			[hep-th]}}.
	
	\bibitem{Bonelli:2018kik}
	G.~Bonelli, N.~Doroud, and M.~Zhu, {\it {$T\bar T$-deformations in closed
			form}},
	\href{http://arxiv.org/abs/1804.10967}{{\ttfamily arXiv:1804.10967 [hep-th]}}.
	
	\bibitem{Henningson:1998gx}
	M.~Henningson and K.~Skenderis, {\it {The Holographic Weyl anomaly}},
	\href{http://dx.doi.org/10.1088/1126-6708/1998/07/023}{{\sf JHEP} {\sf {07}
		}{\sf (1998) }{\sf 023}},
	\href{http://arxiv.org/abs/hep-th/9806087}{{\ttfamily arXiv:hep-th/9806087
			[hep-th]}}.
	
	\bibitem{Ita:2017uvz}
	E.~E. Ita, C.~Soo, and H.-L. Yu, {\it {Intrinsic time gravity, heat kernel
			regularization, and emergence of Einstein's theory}},
	\href{http://arxiv.org/abs/1707.02720}{{\ttfamily arXiv:1707.02720 [gr-qc]}}.
	
	\bibitem{Caputa:2018asc}
	P.~Caputa and S.~Hirano, {\it {Airy Function and 4d Quantum Gravity}},
	\href{http://dx.doi.org/10.1007/JHEP06(2018)106}{{\sf JHEP} {\sf {06} }{\sf
			(2018) }{\sf 106}},
	\href{http://arxiv.org/abs/1804.00942}{{\ttfamily arXiv:1804.00942 [hep-th]}}.
	
	\bibitem{Ryu:2006bv}
	S.~Ryu and T.~Takayanagi, {\it {Holographic derivation of entanglement entropy
			from AdS/CFT}},  \href{http://dx.doi.org/10.1103/PhysRevLett.96.181602}{{\sf
			Phys. Rev. Lett.} {\sf {96} }{\sf (2006) }{\sf 181602}},
	\href{http://arxiv.org/abs/hep-th/0603001}{{\ttfamily arXiv:hep-th/0603001
			[hep-th]}}.
	
	\bibitem{CaputaHiranoWIP}
	P.~Caputa and S.~Hirano, {\it {Work in progress}}, .
	
	\bibitem{deBoer:2000cz}
	J.~de~Boer, {\it {The Holographic renormalization group}},
	\href{http://dx.doi.org/10.1002/1521-3978(200105)49:4/6<339::AID-PROP339>3.0.CO;2-A}{{\sf
			Fortsch. Phys.} {\sf {49} }{\sf (2001) }{\sf 339--358}},
	\href{http://arxiv.org/abs/hep-th/0101026}{{\ttfamily arXiv:hep-th/0101026
			[hep-th]}}.
	
	\bibitem{Papadimitriou:2004ap}
	I.~Papadimitriou and K.~Skenderis, {\it {AdS / CFT correspondence and
			geometry}},  \href{http://dx.doi.org/10.4171/013-1/4}{{\sf IRMA Lect. Math.
			Theor. Phys.} {\sf {8} }{\sf (2005) }{\sf 73--101}},
	\href{http://arxiv.org/abs/hep-th/0404176}{{\ttfamily arXiv:hep-th/0404176
			[hep-th]}}.
	
	\bibitem{Freidel:2008sh}
	L.~Freidel, {\it {Reconstructing AdS/CFT}},
	\href{http://arxiv.org/abs/0804.0632}{{\ttfamily arXiv:0804.0632 [hep-th]}}.
	
	\bibitem{vanRees:2011ir} 
          B.~C.~van Rees, {\it {Irrelevant deformations and the holographic Callan-Symanzik equation}},
           \href{http://link.springer.com/10.1007/JHEP10(2011)067}{{\sf JHEP} {\sf {1110} }{\sf
			(2011) }{\sf 067}}, \href{http://arxiv.org/abs/1105.5396}{{\ttfamily arXiv:1105.5396 [hep-th]}}.
	
	\bibitem{Schwimmer:2019efk}
	A.~Schwimmer and S.~Theisen, {\it {Osborn Equation and Irrelevant Operators}},
	\href{http://arxiv.org/abs/1902.04473}{{\ttfamily arXiv:1902.04473 [hep-th]}}.
	
	\bibitem{Akhavan:2018wla} 
        A.~Akhavan, M.~Alishahiha, A.~Naseh and H.~Zolfi, {\it {Complexity and Behind the Horizon Cut-Off}},
        \href{https://link.springer.com/article/10.1007%2FJHEP12%282018%29090}{{\sf JHEP} {\sf {1812}
	}{\sf (2018) }{\sf 090}}, \href{https://arxiv.org/abs/1810.12015}{{\ttfamily arXiv:1810.12015 [hep-th]}}.
        
        \bibitem{Hashemi:2019xeq} 
       S.~S.~Hashemi, G.~Jafari, A.~Naseh and H.~Zolfi, {\it {More on Complexity in Finite Cut Off Geometry}},      
       \href{https://arxiv.org/abs/1902.03554}{{\ttfamily  arXiv:1902.03554 [hep-th]}}.

	
\end{thebibliography}\endgroup

\providecommand{\href}[2]{#2}

\end{document}